\documentclass[twocolumn,aps,prl,showpacs,superscriptaddress,notitlepage]{revtex4-1}
\usepackage[a4paper,top=3cm,bottom=3cm,left=1.cm,right=1.cm]{geometry}
\usepackage[english]{babel}
\usepackage{graphicx}				
\usepackage{booktabs}			
\usepackage{amsmath}				
\usepackage{mathtools}				
\usepackage{braket}					
\usepackage{multirow}
\usepackage{setspace}
\usepackage[separate-uncertainty=true]{siunitx}
\usepackage{xcolor}
\usepackage{hyperref}
\usepackage{enumitem}

\DeclarePairedDelimiter\abs{\lvert}{\rvert}

\newcommand{\oos}{\omega_s}
\newcommand{\ooi}{\omega_i}

\newcommand{\ooa}{\left( \oos,\ooi\right)}

\makeatletter
\let\cat@comma@active\@empty
\makeatother

\begin{document}
\title{Independent high-purity photons created in domain-engineered crystals}

\author{Francesco Graffitti}
\email{fraccalo@gmail.com} 
\affiliation{Scottish Universities Physics Alliance (SUPA), Institute of Photonics and Quantum Sciences, School of Engineering and Physical Sciences, Heriot-Watt University, Edinburgh EH14 4AS, UK}

\author{Peter Barrow}
\affiliation{Scottish Universities Physics Alliance (SUPA), Institute of Photonics and Quantum Sciences, School of Engineering and Physical Sciences, Heriot-Watt University, Edinburgh EH14 4AS, UK}

\author{Massimiliano Proietti}
\affiliation{Scottish Universities Physics Alliance (SUPA), Institute of Photonics and Quantum Sciences, School of Engineering and Physical Sciences, Heriot-Watt University, Edinburgh EH14 4AS, UK}

\author{Dmytro Kundys}
\affiliation{Scottish Universities Physics Alliance (SUPA), Institute of Photonics and Quantum Sciences, School of Engineering and Physical Sciences, Heriot-Watt University, Edinburgh EH14 4AS, UK}

\author{Alessandro Fedrizzi}
\affiliation{Scottish Universities Physics Alliance (SUPA), Institute of Photonics and Quantum Sciences, School of Engineering and Physical Sciences, Heriot-Watt University, Edinburgh EH14 4AS, UK}

\begin{abstract}
Advanced photonic quantum technology relies on multi-photon interference which requires bright sources of high-purity single photons.
Here, we implement a novel domain-engineering technique for tailoring the nonlinearity of a parametric down-conversion crystal.
We create pairs of independently-heralded telecom-wavelength photons and achieve high heralding, brightness and spectral purities without filtering.
\end{abstract}

\maketitle

The ability of generating and manipulating single quanta of light enables the possibility of exploring new quantum-enhanced technologies. 
Thanks to their robust coherence and the possibility of travelling over long distances with low losses, single photons are the ideal carriers of quantum information in large scale quantum networks \cite{collins2016experimental,sun2017entanglement,liao2017satellite}.
Recent works have shown how photonic platforms are suitable not only for communication purposes, but also play a role in other areas of quantum information, such as quantum computing  \cite{nutz2017efficient,barz2012demonstration,schwartz2016deterministic} and simulation \cite{ma2014towards,wang2017experimental}.
In particular, scalable/fault-tolerant linear optical quantum computing (LOQC) appears to be a promising platform for quantum computing \cite{gimeno2015three,rudolph2017optimistic}. However, LOQC requires photons with near-unity purities and heralding efficiencies \cite{li2010fault,meyer2017limits}, as each percentage point in unsuccessful gate operation or photon loss comes at a significant overhead cost in the required number of photon sources, detectors and circuit complexity \cite{christ2012limits,li2015resource}.

A wide range of single-photon emitters are under development, typically classed into single-quantum emitters such as quantum dots \cite{senellart2017high} and parametric optical processes. The quality of photons and brightness of quantum dot sources is ever increasing, however in many cases parametric downconversion (PDC) in nonlinear crystals still provides a simpler, higher-quality solution especially at telecommunication wavelengths. 
Importantly, despite its probabilistic nature, PDC sources can approximate deterministic on-demand operation in multiplexed schemes \cite{broome2011reducing,collins2013integrated,xiong2016active,kaneda2017quantum} with small overheads.

\begin{figure}[!tb]\center
\includegraphics[width=1\columnwidth]{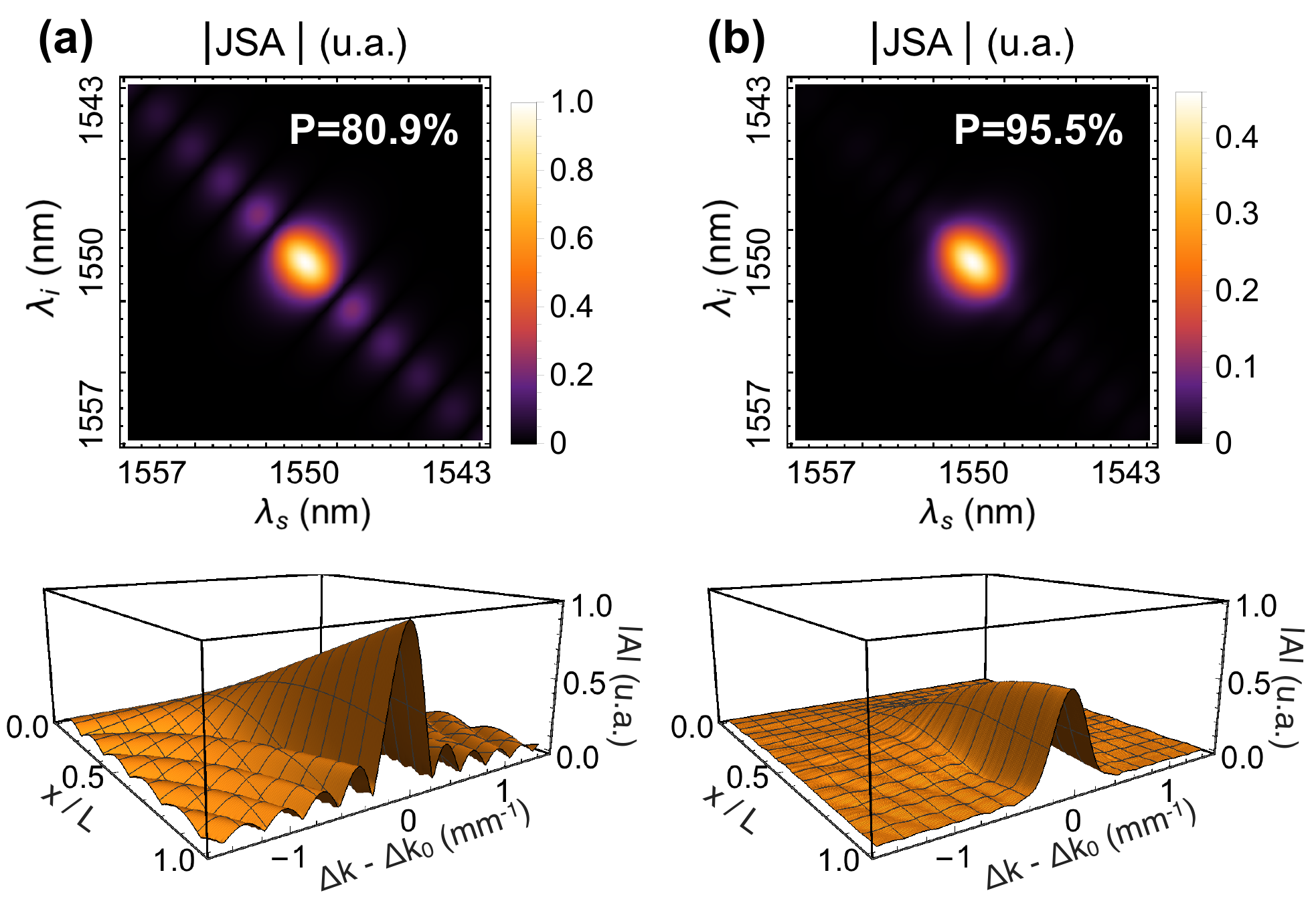}
\caption{
Simulated JSA (top) and field amplitude (bottom) along the crystal for the ppKTP (a) and aKTP (b) pumped with a \SI{1.7}{\pico\second}-\textit{sech$^2$} pulse. The amplitude at the end of the crystal (x=L) corresponds to the PMF, \textit{sinc}-shaped for the standard crystal and \textit{Gaussian} for the apodized one.
The $\Delta k$ dependence on the signal-idler frequencies in the PMF is based on Sellmeier equations reported in \cite{fradkin1999tunable,emanueli2003temperature,konig2004extended}. The heralded-photon spectral purity is computed by performing a numerical Schmidt decomposition on the discretised JSA \cite{laudenbach2016modelling}.
}
\label{JSAcrystals}
\end{figure}

A central requirement for producing high purity heralded photons via PDC is to remove the spectral correlations that naturally arise due to energy conservation: typically, narrowband filters are employed for this purpose at the expense of brightness and heralding efficiency of the source.
This problem can be overcome with a technique known under the umbrella of 'group-velocity matching' \cite{grice2001eliminating,u2006generation,mosley2008conditional}, which involves three separate steps: (i) the averaged PDC photons' group velocities need to be matched to the pump group velocity; (ii) the pump bandwidth needs to match the PDC bandwidth; (iii) the longitudinal nonlinearity profile of the PDC crystal needs to be \textit{Gaussian} in order to avoid spurious correlations arising from the \textit{sinc}-shaped phase matching function (PMF) produced by a standard rectangular crystal nonlinearity \cite{branczyk2011engineered} (see fig. \ref{JSAcrystals}).

Techniques for shaping the effective nonlinearity of poled crystals had long been in use for classical nonlinear optics but have only recently been investigated for the creation of spectrally pure photons \cite{branczyk2011engineered,dixon2013spectral,dosseva2016shaping,tambasco2016domain,graffitti2017pure}. 
Previous proof-of-principle demonstrations verified that domain-engineered Gaussian-profile crystals could indeed be used to create photons with approximated Gaussian spectra~\cite{branczyk2011engineered,dixon2013spectral,chen2017efficient}, and such 'apodized' crystals have already found applications in significant experiments such as a loophole-free Bell test \cite{shalm2015strong}. However, a demonstration of two-photon interference of independently created photons in fully group-velocity matched sources has so far been elusive. 

Here we present a single-photon source that achieves both high signal-idler indistinguishability and high heralded single-photon spectral purity at the same time, without employing narrowband spectral filtering.
We show for the first time that shaping the nonlinearity of a crystal via domain-engineering techniques allows to achieve high visibilities in interference experiments between heralded photons created in two independent PDC processes without the need of additional filtering.

We first consider the first-order PDC bi-photon state:
\begin{equation}
\ket{\psi_\text{pair}}_{s,i} = \iint d\oos d\ooi f\ooa \hat{a}^\dagger_s(\oos) \hat{a}^\dagger_i(\ooi) \ket{0}_{s,i} \, ,\label{PDCspectral}
\end{equation}
where $s$ ($i$) denotes the signal (idler) photon. The joint spectral amplitude (JSA) $f\ooa$, which fully contains the spectral information of the PDC photons, 
depends on the spectral properties of the pump and the phase-matching function (PMF) \cite{grice2001eliminating}, which, in turn, reflects the nonlinear properties of the crystal.
In a heralded single-photon source, where the detection of one photon heralds the presence of another, the spectral purity of the single photon decreases as the spectral correlations between the signal and idler increase \cite{u2006generation}. The JSA therefore has to be separable in order to generate pure photons. This can be achieved by matching the group-velocity of the pump to the average group velocity of signal and idler \cite{grice2001eliminating,u2006generation,mosley2008conditional}, however most experiments still required narrowband (and thus lossy) filtering to remove correlations resulting from rectangular nonlinearity profiles of standard crystals.

The purity and indistinguishability of single photons produced via PDC processes can be determined from the JSA. The JSA is not easily accessible, but one can extract some information from measuring the joint spectral intensity (JSI), e.g. with a pair of standard grating spectrometers in coincidence, or more elegantly with dispersion spectroscopy
\cite{avenhaus2009fiber,weston2016efficient,chen2017efficient}, or via stimulated emission tomography \cite{liscidini2013stimulated,eckstein2014high}.

However, the utility of these measurements is somewhat limited.
Dispersion spectroscopy in long optical fibres shows a significant trade-off between wavelength range, spectral resolution and signal-to-noise ratio due to fibre losses. Recent demonstrations (e.g. \cite{weston2016efficient,chen2017efficient}) focus on the main lobe of the JSI in return for increased spectral resolution, ignoring the significant correlations in the side lobes of the PMF, which have to be taken into account for a reliable purity estimation.
To illustrate this for the standard ppKTP in fig. \ref{JSAcrystals}, restricting the JSA to just one fifth of the range we used in our simulation (i.e. focusing on the main peak, such as in \cite{weston2016efficient,chen2017efficient}) would increase the apparent purity calculated via a discretized Schmidt decomposition \cite{law2000continuous} from \SI{80.9}{\percent} to \SI{93.2}{\percent}.
Furthermore, the JSI doesn't capture the phase information in the JSA, leading to discrepancies between the inferred and the actual single photon purity. 
Performing the Schmidt decomposition on the square root of the JSI instead of the JSA for same crystal yields a spectral purity of \SI{83.9}{\percent} instead of \SI{80.9}{\percent}.
While the first problem can be overcome using stimulated emission tomography, the second one is intrinsic to the measurement of the JSI.

Therefore, a more reasonable benchmark for the single-photon purity and indistinguishability is direct observation via two-photon interference.
Two-photon interference between photons produced by the same PDC process gives an estimate of signal-idler indistinguishability \cite{giovannetti2002extended}. More importantly, the two-photon interference between heralded photons corresponds to the single-photon purity \cite{branczyk2017hong}, where the visibility is defined as the free parameter in the dip fit or, equivalently, as $(N_{\text{max}}-N_{\text{min}})/N_{\text{max}}$ (where $N_{\text{max}}$ ($N_{\text{min}}$) is the maximum (minimum) number of fourfold coincidences).

In order to generate pure photons without compromising the heralding of the source, we design an apodized potassium titanyl phosphate (aKTP) crystal using the domain-engineering annealing-algorithm introduced in \cite{graffitti2017pure} and we compare its performances with a standard periodically-poled KTP (ppKTP).
Starting from a seed poling period of \SI{46.22}{\micro\meter}, our algorithm chooses each ferroelectric domain's orientation and width in order to shape the overall crystal PMF as a \textit{Gaussian} function (fig. \ref{JSAcrystals} (b)). The custom crystal was manufactured by Raicol Crystals Ltd.
KTP crystals are perfectly group-velocity matched for degenerate/collinear type-II PDC with a pump pulse having a \emph{Gaussian} envelope at a wavelength of \SI{791}{\nano\meter} \cite{u2006generation}. 

In fig. \ref{JSAcrystals} we compare the simulated JSAs under our experimental conditions for the aKTP and ppKTP. Taking into account that our pump laser operates at \SI{775}{\nano\metre}, and that a realistic mode-locked laser pulse has a \emph{sech}$^2$ shape, the maximum achievable purity decreases from \SI{98.1}{\percent} to \SI{97.0}{\percent}. The aKTP crystal length is \SI{29}{\milli\meter} while the ppKTP is \SI{22}{\milli\metre} long, so that the corresponding PMF full-width half-maximum (FWHM) of both crystals is matched to a \textit{sech$^2$} transform-limited pulse of \SI{1.4}{\pico\second} duration, where we define the pulse duration as the FWHM of the pulse's intensity envelope. However, our laser isn't stable below \SI{1.7}{\pico\second} (measured with an autocorrelator), and the resulting JSA is slightly elliptical. Under these conditions, we estimate single-photon purities of \SI{80.9}{\percent} and \SI{95.5}{\percent} for the standard and engineered crystals, respectively. These values upper-bound the experimentally achievable two-photon interference visibilities for independently heralded photons.

\begin{figure}[htb]\center
\includegraphics[width=1\columnwidth]{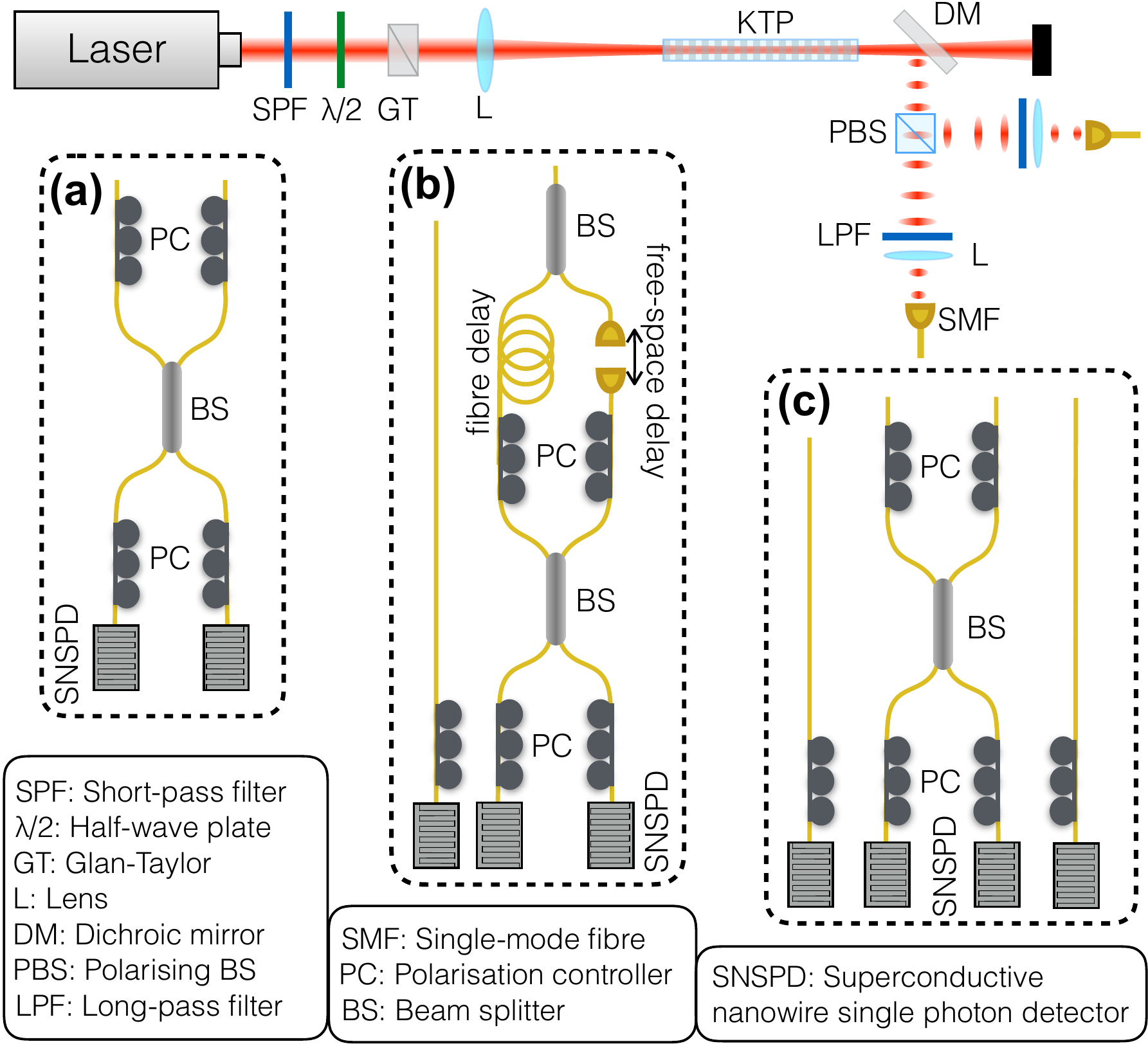}
\caption{
Schematic of the experimental setup. 
(a) Two-photon interference between photons produced by the same PDC process. 
(b) Two-photon interference between photons produced by two different PDC processes by the same crystal at different times.
(c) Two-photon interference between photons produced by two different PDC processes in two different crystals. 
}
\label{setup}
\end{figure}

A description of our experiment is shown in fig. \ref{setup}.
The pump provided by a \SI{80}{\mega\hertz} repetition rate Ti-Sapphire laser first passes through a short-pass filter to suppress any pump superluminescence at telecom band, a Glan-Taylor to ensure the correct polarisation, and then focused in the crystal. 
A dichroic mirror separates the pump from the \SI{1550}{\nano\meter} signal and idler, 
that are successively separated with a PBS, filtered with a long-pass filter for removing any residual from the pump, collected in single-mode fibres and are used in three different setups \ref{setup} (a-c). The photons are finally detected by superconductive nanowire single-photon detectors (SNSPDs).
We measure a source brightness of \SI{11.25\pm0.08 e3}{cc/\milli\watt} and \SI{4.02\pm0.04 e3}{cc/\milli\watt} for the ppKTP and the aKTP, respectively, and a fourfold rate for two independent sources of 
\SI{1.52\pm0.02}{cc/\milli\watt^2} and \SI{0.19\pm0.01}{cc/\milli\watt^2}.
Considering both the singles counts detected by the SNSPDs (s$_1$ and s$_2$) and the corresponding coincidences (cc) recorded by our counting logic within a \SI{1}{\nano\second} time window, we estimate a symmetric heralding efficiency, defined as $\eta = \text{cc}/ (\sqrt{\text{s}_1\text{s}_2})$, of \SI{53}{\percent} in the configuration used for the experiment.
In this configuration, we achieve a two-photon indistinguishability of \SI{98.7\pm 0.3}{\percent} and a heralded-photon spectral-purity of \SI{90.7\pm 0.2}{\percent} without any spectral filtering.
We also find that a heralding efficiency up to \SI{65}{\percent} can be achieved with the same crystals under loose focusing conditions~\cite{shalm2015strong}, at the expense of brightness: this corresponds to a collection efficiency of \SI{88.5}{\percent} once detector efficiency (\SI{80}{\percent}) and known optical losses of (\SI{7.6}{\percent}) are accounted for.

In \eqref{PDCspectral} we considered the spectral properties of the PDC source neglecting higher-order photon-pair emissions. If we omit the spectral contribution to the global state and we look at the system in the Fock space, the PDC state can be written as:
\begin{equation}
\ket{\psi_{\text{PDC}}}=\sqrt{1-\abs{\lambda}^2} \sum_{n=0}^\infty \lambda^n \ket{n}_s \ket{n}_i\, ,
\label{PDCnumber}
\end{equation}
where $n$ is the photon-number. The parameter $\lambda$ is related to the brightness of the source \cite{kok2007linear,broome2011reducing} and can be expressed as a function of the pump power $P$ and of the constant $\tau$, determined by the overall efficiency of the nonlinear process, the detection efficiency and losses in the setup: $\lambda = \sqrt{P \tau}$.
Since perfect two-photon interference occurs when two and only two identical photons enter the 50-50 BS, all terms proportional to $\ket{n>1}_s \ket{n>1}_i$ in eq. \eqref{PDCnumber} compromise the interference visibility.
In the limit of low pump-power, the two-photon visibility decreases linearly with increasing pump power (see appendix A), and we can extrapolate a visibility V$_0$ from a range of measurements at different pump power (i.e for a range of $\lambda \to 0$).
V$_0$ constitutes a lower bound for the indistinguishability (in the case of signal-idler interference) and single-photon spectral purity (in the case of two interfering heralded photons).
 
As a first characterisation of our source, we estimate the signal-idler indistinguishability 
by interfering photons produced by the same PDC processes (see fig.\ref{setup} (a)) at different pump powers. 
\begin{figure*}[t!]\center
\includegraphics[width=1\textwidth]{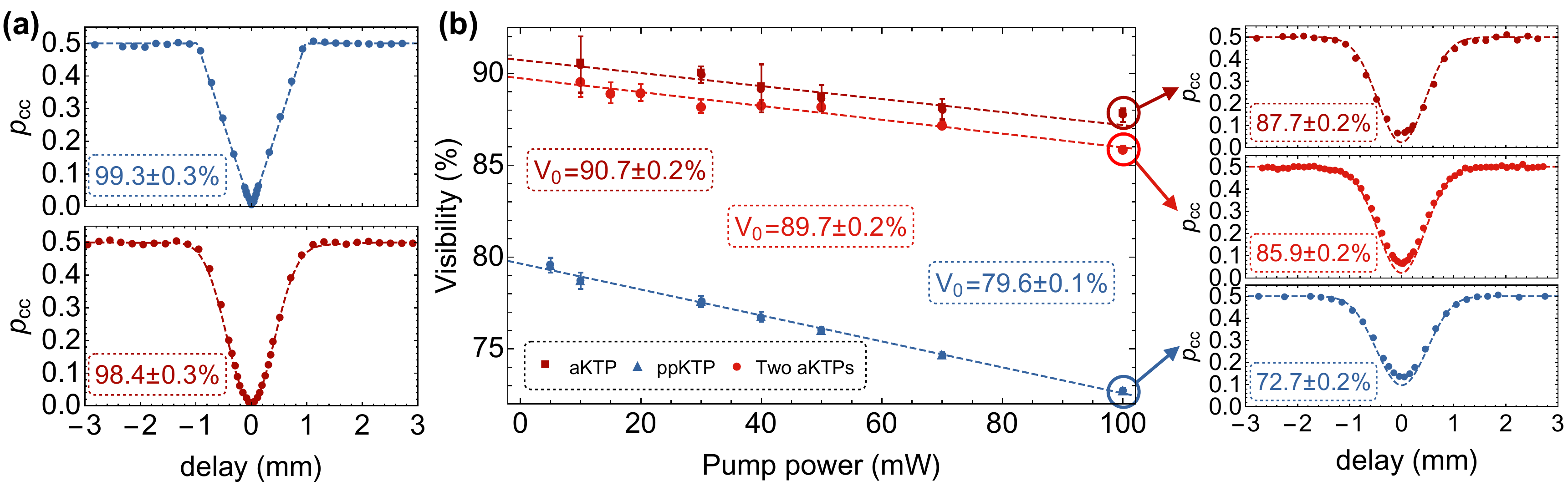}
\caption{
(a) Two-photon interference as a function of temporal delay between photons generated by the same PDC process in the ppKTP (top) and the apodized crystal (bottom), pumped at \SI{20}{\milli\watt}. Data is normalised against a coincidence probability of $1/2$ outside the interference region.
(b) Interference visibilities at different pump powers for the two experimental schemes in fig. \ref{setup} (b,c).
The dashed lines are the simulated interference pattern obtained from the JSA in fig. \ref{JSAcrystals}, while the visibilities are obtained from Gaussian fits. 
Errors are estimated through Monte Carlo simulation based on Poissonian count statistics.}
\label{2fold4foldHOM}
\end{figure*}
Fig. \ref{2fold4foldHOM} (a) shows two-photon interference patterns for the ppKTP and the aKTP at low pump power (\SI{20}{\milli\watt}) without spectral filtering, apart from a long-pass filter at \SI{1319}{\nano\meter} to block residual pump light.
Being in symmetric GVM condition, the two-photon interference pattern can be approximated by the convolution of the phase matching function \cite{barbieri2017hong}: as expected, the two-photon interference pattern is almost \textit{triangular} for the standard crystal \cite{fedrizzi2009anti}, while it is \textit{Gaussian} for the custom design \cite{branczyk2011engineered}. 
We find a signal-idler indistinguishability V$_0$ of $\SI{99.7\pm 0.1}{\percent}$ for the periodically-poled crystal and $\SI{98.7\pm 0.3}{\percent}$ for the apodized crystal which is, to our knowledge, the highest recorded so far with an apodized crystal.

Since purity is quadratic in the density matrix, multiple copies of the same system are required for measuring it. For a PDC photon, that means ideally two photons generated in the same crystal but separated in time.
In our setup (fig.\ref{setup} (b)), this translates into sending the first heralded-photon (generated by the first pump pulse) in a fibre delay line and the second one (generated by the fifth pump pulse) into a shorter fibre, and the two photons are then superposed on a fibre beamsplitter. This process has a success probability of $1/4$, and the pulse delay was chosen to exceed the SNSPD reset time of about \SI{60}{\nano\second}.
The two interfering photons are heralded by their respective twins, and fourfold coincidences are recorded.
We measure a $V_0$ of \SI{79.6\pm0.1}{\%} for the standard ppKTP which matches theory expectations (see figure 1), and we achieve a $V_0$ of \SI{90.7\pm0.2}{\%} with the apodized crystals, see fig. \ref{2fold4foldHOM} (b). 

To show how our technique is feasible for multi-photon experiments, we then interfere and detect photons produced by two different aKTP crystals (figure \ref{setup} (c)). We also detect the idler photons and collect the overall number of fourfold coincidences from the four SNSPDs. In this configuration, we measure a $V_0$ of \SI{89.7\pm0.2}{\%} (fig. \ref{2fold4foldHOM} (b)).

In order to increase indistinguishability and spectral purity of the photons produced with the aKTP, we apply a ``gentle'' spectral filtering. We use a bandpass filter with a spectral transmission function of the form: $e^{-\frac{(\omega-\omega_0)^4}{2 \sigma^4}}$, a central wavelength $\omega_0=$\SI{1550}{\nano\meter} and a FWHM of \SI{7.4}{\nano\meter}, which is roughly five times larger than our PDC photons bandwidth (defined as the FWHM of the marginal spectral distributions of both the idler and the signal).
Such loose filtering doesn't effect sensibly the source heralding efficiency---we measure a decrease of \SI{1}{\percent}--- and in this configuration we achieve a heralded-photon purity of \SI{92.7\pm0.2}{\percent}
and a signal-idler indistinguishability of $\SI{99.7\pm 0.3}{\percent}$, as shown in blue in fig. \ref{2fold4foldHOM} (a): this value is close to the maximum visibility we can achieve ($\SI{99.8}{\percent}$) due to the imperfect fibre BS (reflectivity of \SI{49.2}{\percent}) and bulk PBS (\SI{0.5}{\percent} leakage of the opposite polarised photon).
A similar situation has been observed in PDC photons produced by different aKTPs \cite{chen2017efficient}, where the authors measured an increase of the photons indistinguishability by applying gentle filtering.

Finally, we experimentally prove that nonlinearity-engineering techniques are suitable for generating polarisation-encoded entanglement by using the apodized crystal in a interferometric scheme, such as the Sagnac source originally proposed in \cite{fedrizzi2007wavelength}. We achieve two-photon polarisation purity (\SI{97.4 \pm 0.1}{\percent}) and concurrence (\SI{97.3 \pm 0.1}{\percent}) comparable with what we get with the standard ppKTP.

The $V_0$ corresponding to the apodized crystals shown in fig. \ref{2fold4foldHOM} (a) are definitely higher than the ones corresponding to the standard KTPs: however, they are still not as high as the purity expected from the simulation shown in fig. \ref{JSAcrystals} due to experimental imperfections.
The fibre BS has a reflectivity (transmissivity) $49.2\%$ ($50.8\%$), and the PBS leaks $0.5\%$ of opposite polarised photons. This decreases the visibility by approximately $0.2\%$ for independent photon sources.
We also quantify the random duty-cycle errors that may occur during the fabrication process \cite{fejer1992quasi,pelc2010influence} by performing a Monte Carlo simulation, as an analytical treatment for crystals with strongly aperiodic poling is not straightforward \cite{phillips2013parametric}. 
We randomly vary each domain's width according to a Gaussian distribution with \SI{1}{\micro\meter} FWHM and for each instance compute the JSA and corresponding photon purity. We find a degradation of the mean single-photon purity of about $0.3\%$, with a final value of $P=$\SI{95.2\pm0.2}{\%}. A further degradation is expected due to spatial wavefront distortion due to the significant length of the custom crystals.
Finally, the imperfect indistinguishability of the signal-idler photons, and its increase under gentle filtering suggests the presence of spectral noise in the PDC generation at large $\Delta k - \Delta k_0$ values, which is not present in the standard ppKTP (where the two-photon indistinguishability is nearly perfect).
A more detailed treatment of noise arising in domain-engineered crystals is beyond the scope of this paper and will be addressed separately \cite{graffitti2018poled}.

\begin{figure}[htb]\center
\includegraphics[width=1\columnwidth]{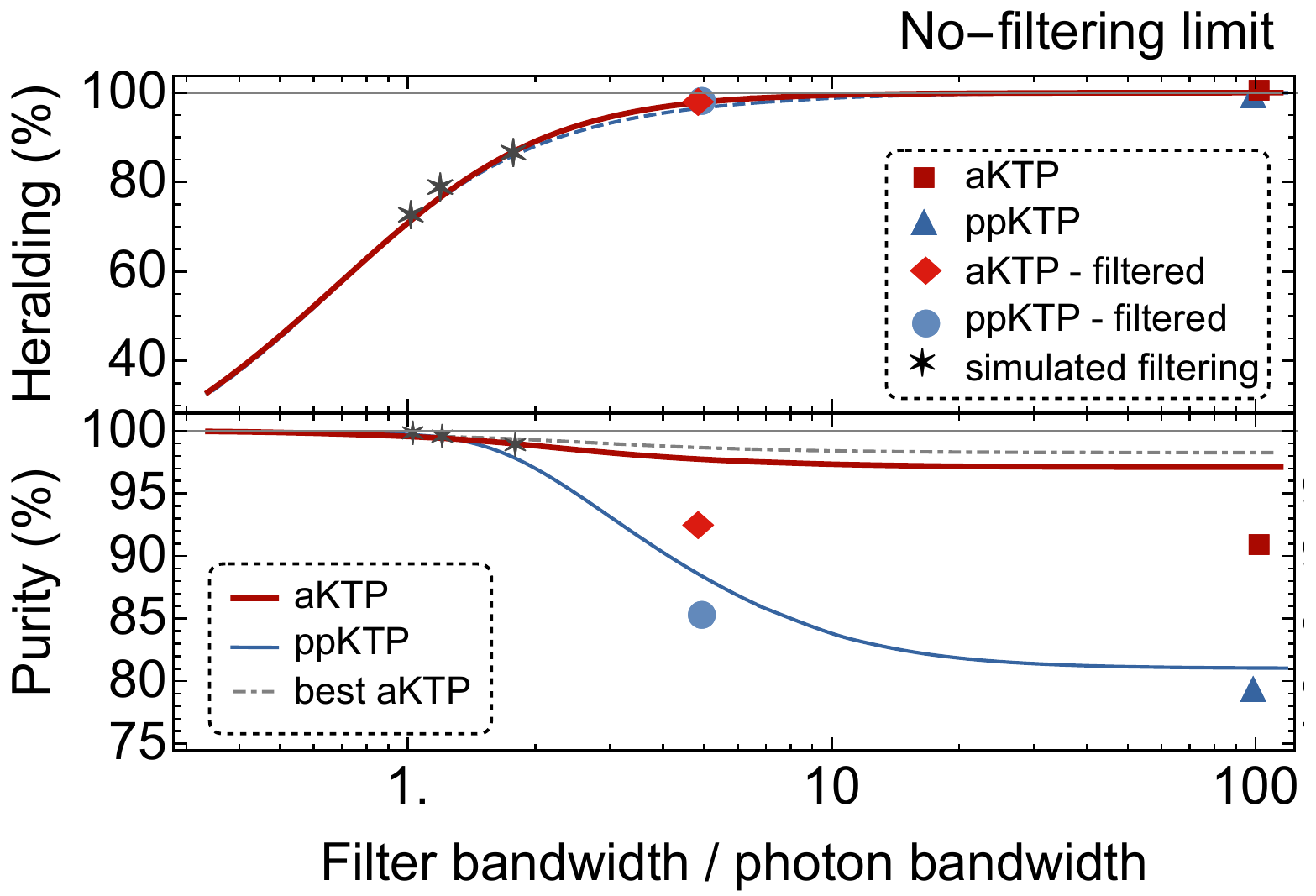}
\caption{
Heralding efficiencies and purities for the ppKTP (blue) and the aKTP (red) under spectral filtering.
The light-blue and light-red data points correspond to the ppKTP and aKTP with gentle filtering, while the dark grey stars are simulation for three commercially available narrowband filters (from left to right: Iridian Spectral Technologies Ltd., Alluxa, Omega Optical Inc.) applied to ppKTP crystal. The dot-dashed grey line shows the simulated purity for a crystal designed with the optimal algorithm described in \cite{graffitti2017pure}.}
\label{filtering}
\end{figure}

To conclude we assess in fig. \ref{filtering} how filtering periodically-poled and apodized crystals affect the heralding efficiency and the single-photon purity of a PDC photon source. The normalized heralding represents the maximum heralding achievable by the source, ruling out losses, detection and collection efficiency, while the x axis is the ratio between the filter bandwidth and the single-photon bandwidth. Note that, while the experimental data corresponds to our setup, the simulated heralding and purities hold in general for ppKTP/aKTP crystals of any length when group velocity matched with a \textit{sech$^2$} pulse. 
The figure shows the significant trade-off between spectral purity and heralding efficiency in the ppKTP case.
For example, a ppKTP can produce single photons with \SI{99}{\percent} purity when a filter having twice the bandwidth of the PDC photons is applied to both signal and idler. However,
even in the case of an ideal filter with \SI{100}{\percent} peak transmission, this would lead to a heralding efficiency of \SI{80}{\percent}, and to a drop of the source brightness to \SI{60}{\percent} of the original pair generation, which consequently translates into a drop of the fourfold and sixfolds events to \SI{36}{\percent} and \SI{22}{\percent}, respectively.
It is then clear that achieving high single-photon purities and, at the same time, maintaining high heralding efficiencies and count rates (especially in many photon experiments), requires efforts that go beyond the mere spectral filtering, and our crystal engineering technique is a suitable solution to this problem.
\newpage

\section*{Funding Information}
Engineering and Physical Sciences Research Council (EPSRC) (EP/N002962/1,EP/L015110/1). 

\section*{Acknowledgments}
The authors thank A. M. Bra\'nczyk for useful discussions and A. Pickston and M. Ringbauer{\tiny} for experimental assistance.

\bibliography{bibliography_purephotons}{}

\begin{thebibliography}{49}%
\makeatletter
\providecommand \@ifxundefined [1]{%
 \@ifx{#1\undefined}
}%
\providecommand \@ifnum [1]{%
 \ifnum #1\expandafter \@firstoftwo
 \else \expandafter \@secondoftwo
 \fi
}%
\providecommand \@ifx [1]{%
 \ifx #1\expandafter \@firstoftwo
 \else \expandafter \@secondoftwo
 \fi
}%
\providecommand \natexlab [1]{#1}%
\providecommand \enquote  [1]{``#1''}%
\providecommand \bibnamefont  [1]{#1}%
\providecommand \bibfnamefont [1]{#1}%
\providecommand \citenamefont [1]{#1}%
\providecommand \href@noop [0]{\@secondoftwo}%
\providecommand \href [0]{\begingroup \@sanitize@url \@href}%
\providecommand \@href[1]{\@@startlink{#1}\@@href}%
\providecommand \@@href[1]{\endgroup#1\@@endlink}%
\providecommand \@sanitize@url [0]{\catcode `\\12\catcode `\$12\catcode
  `\&12\catcode `\#12\catcode `\^12\catcode `\_12\catcode `\%12\relax}%
\providecommand \@@startlink[1]{}%
\providecommand \@@endlink[0]{}%
\providecommand \url  [0]{\begingroup\@sanitize@url \@url }%
\providecommand \@url [1]{\endgroup\@href {#1}{\urlprefix }}%
\providecommand \urlprefix  [0]{URL }%
\providecommand \Eprint [0]{\href }%
\providecommand \doibase [0]{http://dx.doi.org/}%
\providecommand \selectlanguage [0]{\@gobble}%
\providecommand \bibinfo  [0]{\@secondoftwo}%
\providecommand \bibfield  [0]{\@secondoftwo}%
\providecommand \translation [1]{[#1]}%
\providecommand \BibitemOpen [0]{}%
\providecommand \bibitemStop [0]{}%
\providecommand \bibitemNoStop [0]{.\EOS\space}%
\providecommand \EOS [0]{\spacefactor3000\relax}%
\providecommand \BibitemShut  [1]{\csname bibitem#1\endcsname}%
\let\auto@bib@innerbib\@empty
\bibitem [{\citenamefont {Collins}\ \emph {et~al.}(2016)\citenamefont
  {Collins}, \citenamefont {Amiri}, \citenamefont {Fujiwara}, \citenamefont
  {Honjo}, \citenamefont {Shimizu}, \citenamefont {Tamaki}, \citenamefont
  {Takeoka}, \citenamefont {Andersson}, \citenamefont {Buller},\ and\
  \citenamefont {Sasaki}}]{collins2016experimental}%
  \BibitemOpen
  \bibfield  {author} {\bibinfo {author} {\bibfnamefont {R.~J.}\ \bibnamefont
  {Collins}}, \bibinfo {author} {\bibfnamefont {R.}~\bibnamefont {Amiri}},
  \bibinfo {author} {\bibfnamefont {M.}~\bibnamefont {Fujiwara}}, \bibinfo
  {author} {\bibfnamefont {T.}~\bibnamefont {Honjo}}, \bibinfo {author}
  {\bibfnamefont {K.}~\bibnamefont {Shimizu}}, \bibinfo {author} {\bibfnamefont
  {K.}~\bibnamefont {Tamaki}}, \bibinfo {author} {\bibfnamefont
  {M.}~\bibnamefont {Takeoka}}, \bibinfo {author} {\bibfnamefont
  {E.}~\bibnamefont {Andersson}}, \bibinfo {author} {\bibfnamefont {G.~S.}\
  \bibnamefont {Buller}}, \ and\ \bibinfo {author} {\bibfnamefont
  {M.}~\bibnamefont {Sasaki}},\ }\href@noop {} {\bibfield  {journal} {\bibinfo
  {journal} {Optics Letters}\ }\textbf {\bibinfo {volume} {41}},\ \bibinfo
  {pages} {4883} (\bibinfo {year} {2016})}\BibitemShut {NoStop}%
\bibitem [{\citenamefont {Sun}\ \emph {et~al.}(2017)\citenamefont {Sun},
  \citenamefont {Jiang}, \citenamefont {Mao}, \citenamefont {You},
  \citenamefont {Zhang}, \citenamefont {Zhang}, \citenamefont {Jiang},
  \citenamefont {Chen}, \citenamefont {Li}, \citenamefont {Huang} \emph
  {et~al.}}]{sun2017entanglement}%
  \BibitemOpen
  \bibfield  {author} {\bibinfo {author} {\bibfnamefont {Q.-C.}\ \bibnamefont
  {Sun}}, \bibinfo {author} {\bibfnamefont {Y.-F.}\ \bibnamefont {Jiang}},
  \bibinfo {author} {\bibfnamefont {Y.-L.}\ \bibnamefont {Mao}}, \bibinfo
  {author} {\bibfnamefont {L.-X.}\ \bibnamefont {You}}, \bibinfo {author}
  {\bibfnamefont {W.}~\bibnamefont {Zhang}}, \bibinfo {author} {\bibfnamefont
  {W.-J.}\ \bibnamefont {Zhang}}, \bibinfo {author} {\bibfnamefont
  {X.}~\bibnamefont {Jiang}}, \bibinfo {author} {\bibfnamefont {T.-Y.}\
  \bibnamefont {Chen}}, \bibinfo {author} {\bibfnamefont {H.}~\bibnamefont
  {Li}}, \bibinfo {author} {\bibfnamefont {Y.-D.}\ \bibnamefont {Huang}},
  \emph {et~al.},\ }\href@noop {} {\bibfield  {journal} {\bibinfo  {journal}
  {Optica}\ }\textbf {\bibinfo {volume} {4}},\ \bibinfo {pages} {1214}
  (\bibinfo {year} {2017})}\BibitemShut {NoStop}%
\bibitem [{\citenamefont {Liao}\ \emph {et~al.}(2017)\citenamefont {Liao},
  \citenamefont {Cai}, \citenamefont {Liu}, \citenamefont {Zhang},
  \citenamefont {Li}, \citenamefont {Ren}, \citenamefont {Yin}, \citenamefont
  {Shen}, \citenamefont {Cao}, \citenamefont {Li} \emph
  {et~al.}}]{liao2017satellite}%
  \BibitemOpen
  \bibfield  {author} {\bibinfo {author} {\bibfnamefont {S.-K.}\ \bibnamefont
  {Liao}}, \bibinfo {author} {\bibfnamefont {W.-Q.}\ \bibnamefont {Cai}},
  \bibinfo {author} {\bibfnamefont {W.-Y.}\ \bibnamefont {Liu}}, \bibinfo
  {author} {\bibfnamefont {L.}~\bibnamefont {Zhang}}, \bibinfo {author}
  {\bibfnamefont {Y.}~\bibnamefont {Li}}, \bibinfo {author} {\bibfnamefont
  {J.-G.}\ \bibnamefont {Ren}}, \bibinfo {author} {\bibfnamefont
  {J.}~\bibnamefont {Yin}}, \bibinfo {author} {\bibfnamefont {Q.}~\bibnamefont
  {Shen}}, \bibinfo {author} {\bibfnamefont {Y.}~\bibnamefont {Cao}}, \bibinfo
  {author} {\bibfnamefont {Z.-P.}\ \bibnamefont {Li}},  \emph {et~al.},\
  }\href@noop {} {\bibfield  {journal} {\bibinfo  {journal} {arXiv preprint
  arXiv:1707.00542}\ } (\bibinfo {year} {2017})}\BibitemShut {NoStop}%
\bibitem [{\citenamefont {Nutz}\ \emph {et~al.}(2017)\citenamefont {Nutz},
  \citenamefont {Milne}, \citenamefont {Shadbolt},\ and\ \citenamefont
  {Rudolph}}]{nutz2017efficient}%
  \BibitemOpen
  \bibfield  {author} {\bibinfo {author} {\bibfnamefont {T.}~\bibnamefont
  {Nutz}}, \bibinfo {author} {\bibfnamefont {A.}~\bibnamefont {Milne}},
  \bibinfo {author} {\bibfnamefont {P.}~\bibnamefont {Shadbolt}}, \ and\
  \bibinfo {author} {\bibfnamefont {T.}~\bibnamefont {Rudolph}},\ }\href
  {\doibase 10.1063/1.4983822} {\bibfield  {journal} {\bibinfo  {journal} {APL
  Photonics}\ }\textbf {\bibinfo {volume} {2}},\ \bibinfo {pages} {066103}
  (\bibinfo {year} {2017})},\ \Eprint
  {http://arxiv.org/abs/http://dx.doi.org/10.1063/1.4983822}
  {http://dx.doi.org/10.1063/1.4983822} \BibitemShut {NoStop}%
\bibitem [{\citenamefont {Barz}\ \emph {et~al.}(2012)\citenamefont {Barz},
  \citenamefont {Kashefi}, \citenamefont {Broadbent}, \citenamefont
  {Fitzsimons}, \citenamefont {Zeilinger},\ and\ \citenamefont
  {Walther}}]{barz2012demonstration}%
  \BibitemOpen
  \bibfield  {author} {\bibinfo {author} {\bibfnamefont {S.}~\bibnamefont
  {Barz}}, \bibinfo {author} {\bibfnamefont {E.}~\bibnamefont {Kashefi}},
  \bibinfo {author} {\bibfnamefont {A.}~\bibnamefont {Broadbent}}, \bibinfo
  {author} {\bibfnamefont {J.~F.}\ \bibnamefont {Fitzsimons}}, \bibinfo
  {author} {\bibfnamefont {A.}~\bibnamefont {Zeilinger}}, \ and\ \bibinfo
  {author} {\bibfnamefont {P.}~\bibnamefont {Walther}},\ }\href@noop {}
  {\bibfield  {journal} {\bibinfo  {journal} {Science}\ }\textbf {\bibinfo
  {volume} {335}},\ \bibinfo {pages} {303} (\bibinfo {year}
  {2012})}\BibitemShut {NoStop}%
\bibitem [{\citenamefont {Schwartz}\ \emph {et~al.}(2016)\citenamefont
  {Schwartz}, \citenamefont {Cogan}, \citenamefont {Schmidgall}, \citenamefont
  {Don}, \citenamefont {Gantz}, \citenamefont {Kenneth}, \citenamefont
  {Lindner},\ and\ \citenamefont {Gershoni}}]{schwartz2016deterministic}%
  \BibitemOpen
  \bibfield  {author} {\bibinfo {author} {\bibfnamefont {I.}~\bibnamefont
  {Schwartz}}, \bibinfo {author} {\bibfnamefont {D.}~\bibnamefont {Cogan}},
  \bibinfo {author} {\bibfnamefont {E.~R.}\ \bibnamefont {Schmidgall}},
  \bibinfo {author} {\bibfnamefont {Y.}~\bibnamefont {Don}}, \bibinfo {author}
  {\bibfnamefont {L.}~\bibnamefont {Gantz}}, \bibinfo {author} {\bibfnamefont
  {O.}~\bibnamefont {Kenneth}}, \bibinfo {author} {\bibfnamefont {N.~H.}\
  \bibnamefont {Lindner}}, \ and\ \bibinfo {author} {\bibfnamefont
  {D.}~\bibnamefont {Gershoni}},\ }\href@noop {} {\bibfield  {journal}
  {\bibinfo  {journal} {Science}\ ,\ \bibinfo {pages} {aah4758}} (\bibinfo
  {year} {2016})}\BibitemShut {NoStop}%
\bibitem [{\citenamefont {Ma}\ \emph {et~al.}(2014)\citenamefont {Ma},
  \citenamefont {Daki{\'c}}, \citenamefont {Kropatschek}, \citenamefont
  {Naylor}, \citenamefont {Chan}, \citenamefont {Gong}, \citenamefont {Duan},
  \citenamefont {Zeilinger},\ and\ \citenamefont {Walther}}]{ma2014towards}%
  \BibitemOpen
  \bibfield  {author} {\bibinfo {author} {\bibfnamefont {X.-S.}\ \bibnamefont
  {Ma}}, \bibinfo {author} {\bibfnamefont {B.}~\bibnamefont {Daki{\'c}}},
  \bibinfo {author} {\bibfnamefont {S.}~\bibnamefont {Kropatschek}}, \bibinfo
  {author} {\bibfnamefont {W.}~\bibnamefont {Naylor}}, \bibinfo {author}
  {\bibfnamefont {Y.-h.}\ \bibnamefont {Chan}}, \bibinfo {author}
  {\bibfnamefont {Z.-x.}\ \bibnamefont {Gong}}, \bibinfo {author}
  {\bibfnamefont {L.-m.}\ \bibnamefont {Duan}}, \bibinfo {author}
  {\bibfnamefont {A.}~\bibnamefont {Zeilinger}}, \ and\ \bibinfo {author}
  {\bibfnamefont {P.}~\bibnamefont {Walther}},\ }\href@noop {} {\bibfield
  {journal} {\bibinfo  {journal} {Scientific Reports}\ }\textbf {\bibinfo
  {volume} {4}},\ \bibinfo {pages} {3583} (\bibinfo {year} {2014})}\BibitemShut
  {NoStop}%
\bibitem [{\citenamefont {Wang}\ \emph {et~al.}(2017)\citenamefont {Wang},
  \citenamefont {Paesani}, \citenamefont {Santagati}, \citenamefont {Knauer},
  \citenamefont {Gentile}, \citenamefont {Wiebe}, \citenamefont {Petruzzella},
  \citenamefont {O'Brien}, \citenamefont {Rarity}, \citenamefont {Laing},\ and\
  \citenamefont {Thompson}}]{wang2017experimental}%
  \BibitemOpen
  \bibfield  {author} {\bibinfo {author} {\bibfnamefont {J.}~\bibnamefont
  {Wang}}, \bibinfo {author} {\bibfnamefont {S.}~\bibnamefont {Paesani}},
  \bibinfo {author} {\bibfnamefont {R.}~\bibnamefont {Santagati}}, \bibinfo
  {author} {\bibfnamefont {S.}~\bibnamefont {Knauer}}, \bibinfo {author}
  {\bibfnamefont {A.~A.}\ \bibnamefont {Gentile}}, \bibinfo {author}
  {\bibfnamefont {N.}~\bibnamefont {Wiebe}}, \bibinfo {author} {\bibfnamefont
  {M.}~\bibnamefont {Petruzzella}}, \bibinfo {author} {\bibfnamefont {J.~L.}\
  \bibnamefont {O'Brien}}, \bibinfo {author} {\bibfnamefont {J.~G.}\
  \bibnamefont {Rarity}}, \bibinfo {author} {\bibfnamefont {A.}~\bibnamefont
  {Laing}}, \ and\ \bibinfo {author} {\bibfnamefont {M.~G.}\ \bibnamefont
  {Thompson}},\ }\href {http://dx.doi.org/10.1038/nphys4074} {\ \textbf
  {\bibinfo {volume} {13}},\ \bibinfo {pages} {551 EP } (\bibinfo {year}
  {2017})}\BibitemShut {NoStop}%
\bibitem [{\citenamefont {Gimeno-Segovia}\ \emph {et~al.}(2015)\citenamefont
  {Gimeno-Segovia}, \citenamefont {Shadbolt}, \citenamefont {Browne},\ and\
  \citenamefont {Rudolph}}]{gimeno2015three}%
  \BibitemOpen
  \bibfield  {author} {\bibinfo {author} {\bibfnamefont {M.}~\bibnamefont
  {Gimeno-Segovia}}, \bibinfo {author} {\bibfnamefont {P.}~\bibnamefont
  {Shadbolt}}, \bibinfo {author} {\bibfnamefont {D.~E.}\ \bibnamefont
  {Browne}}, \ and\ \bibinfo {author} {\bibfnamefont {T.}~\bibnamefont
  {Rudolph}},\ }\href@noop {} {\bibfield  {journal} {\bibinfo  {journal}
  {Physical Review Letters}\ }\textbf {\bibinfo {volume} {115}},\ \bibinfo
  {pages} {020502} (\bibinfo {year} {2015})}\BibitemShut {NoStop}%
\bibitem [{\citenamefont {Rudolph}(2017)}]{rudolph2017optimistic}%
  \BibitemOpen
  \bibfield  {author} {\bibinfo {author} {\bibfnamefont {T.}~\bibnamefont
  {Rudolph}},\ }\href@noop {} {\bibfield  {journal} {\bibinfo  {journal} {APL
  Photonics}\ }\textbf {\bibinfo {volume} {2}},\ \bibinfo {pages} {030901}
  (\bibinfo {year} {2017})}\BibitemShut {NoStop}%
\bibitem [{\citenamefont {Li}\ \emph {et~al.}(2010)\citenamefont {Li},
  \citenamefont {Barrett}, \citenamefont {Stace},\ and\ \citenamefont
  {Benjamin}}]{li2010fault}%
  \BibitemOpen
  \bibfield  {author} {\bibinfo {author} {\bibfnamefont {Y.}~\bibnamefont
  {Li}}, \bibinfo {author} {\bibfnamefont {S.~D.}\ \bibnamefont {Barrett}},
  \bibinfo {author} {\bibfnamefont {T.~M.}\ \bibnamefont {Stace}}, \ and\
  \bibinfo {author} {\bibfnamefont {S.~C.}\ \bibnamefont {Benjamin}},\
  }\href@noop {} {\bibfield  {journal} {\bibinfo  {journal} {Physical Review
  Letters}\ }\textbf {\bibinfo {volume} {105}},\ \bibinfo {pages} {250502}
  (\bibinfo {year} {2010})}\BibitemShut {NoStop}%
\bibitem [{\citenamefont {Meyer-Scott}\ \emph {et~al.}(2017)\citenamefont
  {Meyer-Scott}, \citenamefont {Montaut}, \citenamefont {Tiedau}, \citenamefont
  {Sansoni}, \citenamefont {Herrmann}, \citenamefont {Bartley},\ and\
  \citenamefont {Silberhorn}}]{meyer2017limits}%
  \BibitemOpen
  \bibfield  {author} {\bibinfo {author} {\bibfnamefont {E.}~\bibnamefont
  {Meyer-Scott}}, \bibinfo {author} {\bibfnamefont {N.}~\bibnamefont
  {Montaut}}, \bibinfo {author} {\bibfnamefont {J.}~\bibnamefont {Tiedau}},
  \bibinfo {author} {\bibfnamefont {L.}~\bibnamefont {Sansoni}}, \bibinfo
  {author} {\bibfnamefont {H.}~\bibnamefont {Herrmann}}, \bibinfo {author}
  {\bibfnamefont {T.~J.}\ \bibnamefont {Bartley}}, \ and\ \bibinfo {author}
  {\bibfnamefont {C.}~\bibnamefont {Silberhorn}},\ }\href@noop {} {\bibfield
  {journal} {\bibinfo  {journal} {Physical Review A}\ }\textbf {\bibinfo
  {volume} {95}},\ \bibinfo {pages} {061803} (\bibinfo {year}
  {2017})}\BibitemShut {NoStop}%
\bibitem [{\citenamefont {Christ}\ and\ \citenamefont
  {Silberhorn}(2012)}]{christ2012limits}%
  \BibitemOpen
  \bibfield  {author} {\bibinfo {author} {\bibfnamefont {A.}~\bibnamefont
  {Christ}}\ and\ \bibinfo {author} {\bibfnamefont {C.}~\bibnamefont
  {Silberhorn}},\ }\href@noop {} {\bibfield  {journal} {\bibinfo  {journal}
  {Physical Review A}\ }\textbf {\bibinfo {volume} {85}},\ \bibinfo {pages}
  {023829} (\bibinfo {year} {2012})}\BibitemShut {NoStop}%
\bibitem [{\citenamefont {Li}\ \emph {et~al.}(2015)\citenamefont {Li},
  \citenamefont {Humphreys}, \citenamefont {Mendoza},\ and\ \citenamefont
  {Benjamin}}]{li2015resource}%
  \BibitemOpen
  \bibfield  {author} {\bibinfo {author} {\bibfnamefont {Y.}~\bibnamefont
  {Li}}, \bibinfo {author} {\bibfnamefont {P.~C.}\ \bibnamefont {Humphreys}},
  \bibinfo {author} {\bibfnamefont {G.~J.}\ \bibnamefont {Mendoza}}, \ and\
  \bibinfo {author} {\bibfnamefont {S.~C.}\ \bibnamefont {Benjamin}},\
  }\href@noop {} {\bibfield  {journal} {\bibinfo  {journal} {Physical Review
  X}\ }\textbf {\bibinfo {volume} {5}},\ \bibinfo {pages} {041007} (\bibinfo
  {year} {2015})}\BibitemShut {NoStop}%
\bibitem [{\citenamefont {Senellart}\ \emph {et~al.}(2017)\citenamefont
  {Senellart}, \citenamefont {Solomon},\ and\ \citenamefont
  {White}}]{senellart2017high}%
  \BibitemOpen
  \bibfield  {author} {\bibinfo {author} {\bibfnamefont {P.}~\bibnamefont
  {Senellart}}, \bibinfo {author} {\bibfnamefont {G.}~\bibnamefont {Solomon}},
  \ and\ \bibinfo {author} {\bibfnamefont {A.}~\bibnamefont {White}},\
  }\href@noop {} {\bibfield  {journal} {\bibinfo  {journal} {Nature
  Nanotechnology}\ }\textbf {\bibinfo {volume} {12}},\ \bibinfo {pages} {nnano}
  (\bibinfo {year} {2017})}\BibitemShut {NoStop}%
\bibitem [{\citenamefont {Broome}\ \emph {et~al.}(2011)\citenamefont {Broome},
  \citenamefont {Almeida}, \citenamefont {Fedrizzi},\ and\ \citenamefont
  {White}}]{broome2011reducing}%
  \BibitemOpen
  \bibfield  {author} {\bibinfo {author} {\bibfnamefont {M.~A.}\ \bibnamefont
  {Broome}}, \bibinfo {author} {\bibfnamefont {M.~P.}\ \bibnamefont {Almeida}},
  \bibinfo {author} {\bibfnamefont {A.}~\bibnamefont {Fedrizzi}}, \ and\
  \bibinfo {author} {\bibfnamefont {A.~G.}\ \bibnamefont {White}},\ }\href@noop
  {} {\bibfield  {journal} {\bibinfo  {journal} {Optics Express}\ }\textbf
  {\bibinfo {volume} {19}},\ \bibinfo {pages} {22698} (\bibinfo {year}
  {2011})}\BibitemShut {NoStop}%
\bibitem [{\citenamefont {Collins}\ \emph {et~al.}(2013)\citenamefont
  {Collins}, \citenamefont {Xiong}, \citenamefont {Rey}, \citenamefont {Vo},
  \citenamefont {He}, \citenamefont {Shahnia}, \citenamefont {Reardon},
  \citenamefont {Krauss}, \citenamefont {Steel}, \citenamefont {Clark} \emph
  {et~al.}}]{collins2013integrated}%
  \BibitemOpen
  \bibfield  {author} {\bibinfo {author} {\bibfnamefont {M.~J.}\ \bibnamefont
  {Collins}}, \bibinfo {author} {\bibfnamefont {C.}~\bibnamefont {Xiong}},
  \bibinfo {author} {\bibfnamefont {I.~H.}\ \bibnamefont {Rey}}, \bibinfo
  {author} {\bibfnamefont {T.~D.}\ \bibnamefont {Vo}}, \bibinfo {author}
  {\bibfnamefont {J.}~\bibnamefont {He}}, \bibinfo {author} {\bibfnamefont
  {S.}~\bibnamefont {Shahnia}}, \bibinfo {author} {\bibfnamefont
  {C.}~\bibnamefont {Reardon}}, \bibinfo {author} {\bibfnamefont {T.~F.}\
  \bibnamefont {Krauss}}, \bibinfo {author} {\bibfnamefont {M.}~\bibnamefont
  {Steel}}, \bibinfo {author} {\bibfnamefont {A.~S.}\ \bibnamefont {Clark}},
  \emph {et~al.},\ }\href@noop {} {\bibfield  {journal} {\bibinfo  {journal}
  {Nature Communications}\ }\textbf {\bibinfo {volume} {4}} (\bibinfo {year}
  {2013})}\BibitemShut {NoStop}%
\bibitem [{\citenamefont {Xiong}\ \emph {et~al.}(2016)\citenamefont {Xiong},
  \citenamefont {Zhang}, \citenamefont {Liu}, \citenamefont {Collins},
  \citenamefont {Mahendra}, \citenamefont {Helt}, \citenamefont {Steel},
  \citenamefont {Choi}, \citenamefont {Chae}, \citenamefont {Leong} \emph
  {et~al.}}]{xiong2016active}%
  \BibitemOpen
  \bibfield  {author} {\bibinfo {author} {\bibfnamefont {C.}~\bibnamefont
  {Xiong}}, \bibinfo {author} {\bibfnamefont {X.}~\bibnamefont {Zhang}},
  \bibinfo {author} {\bibfnamefont {Z.}~\bibnamefont {Liu}}, \bibinfo {author}
  {\bibfnamefont {M.}~\bibnamefont {Collins}}, \bibinfo {author} {\bibfnamefont
  {A.}~\bibnamefont {Mahendra}}, \bibinfo {author} {\bibfnamefont
  {L.}~\bibnamefont {Helt}}, \bibinfo {author} {\bibfnamefont {M.}~\bibnamefont
  {Steel}}, \bibinfo {author} {\bibfnamefont {D.-Y.}\ \bibnamefont {Choi}},
  \bibinfo {author} {\bibfnamefont {C.}~\bibnamefont {Chae}}, \bibinfo {author}
  {\bibfnamefont {P.}~\bibnamefont {Leong}},  \emph {et~al.},\ }\href@noop {}
  {\bibfield  {journal} {\bibinfo  {journal} {Nature Communications}\ }\textbf
  {\bibinfo {volume} {7}} (\bibinfo {year} {2016})}\BibitemShut {NoStop}%
\bibitem [{\citenamefont {Kaneda}\ \emph {et~al.}(2017)\citenamefont {Kaneda},
  \citenamefont {Xu}, \citenamefont {Chapman},\ and\ \citenamefont
  {Kwiat}}]{kaneda2017quantum}%
  \BibitemOpen
  \bibfield  {author} {\bibinfo {author} {\bibfnamefont {F.}~\bibnamefont
  {Kaneda}}, \bibinfo {author} {\bibfnamefont {F.}~\bibnamefont {Xu}}, \bibinfo
  {author} {\bibfnamefont {J.}~\bibnamefont {Chapman}}, \ and\ \bibinfo
  {author} {\bibfnamefont {P.~G.}\ \bibnamefont {Kwiat}},\ }\href@noop {}
  {\bibfield  {journal} {\bibinfo  {journal} {Optica}\ }\textbf {\bibinfo
  {volume} {4}},\ \bibinfo {pages} {1034} (\bibinfo {year} {2017})}\BibitemShut
  {NoStop}%
\bibitem [{\citenamefont {Grice}\ \emph {et~al.}(2001)\citenamefont {Grice},
  \citenamefont {U'Ren},\ and\ \citenamefont
  {Walmsley}}]{grice2001eliminating}%
  \BibitemOpen
  \bibfield  {author} {\bibinfo {author} {\bibfnamefont {W.}~\bibnamefont
  {Grice}}, \bibinfo {author} {\bibfnamefont {A.}~\bibnamefont {U'Ren}}, \ and\
  \bibinfo {author} {\bibfnamefont {I.}~\bibnamefont {Walmsley}},\ }\href@noop
  {} {\bibfield  {journal} {\bibinfo  {journal} {Physical Review A}\ }\textbf
  {\bibinfo {volume} {64}},\ \bibinfo {pages} {063815} (\bibinfo {year}
  {2001})}\BibitemShut {NoStop}%
\bibitem [{\citenamefont {U'Ren}\ \emph {et~al.}(2006)\citenamefont {U'Ren},
  \citenamefont {Silberhorn}, \citenamefont {Erdmann}, \citenamefont
  {Banaszek}, \citenamefont {Grice}, \citenamefont {Walmsley},\ and\
  \citenamefont {Raymer}}]{u2006generation}%
  \BibitemOpen
  \bibfield  {author} {\bibinfo {author} {\bibfnamefont {A.~B.}\ \bibnamefont
  {U'Ren}}, \bibinfo {author} {\bibfnamefont {C.}~\bibnamefont {Silberhorn}},
  \bibinfo {author} {\bibfnamefont {R.}~\bibnamefont {Erdmann}}, \bibinfo
  {author} {\bibfnamefont {K.}~\bibnamefont {Banaszek}}, \bibinfo {author}
  {\bibfnamefont {W.~P.}\ \bibnamefont {Grice}}, \bibinfo {author}
  {\bibfnamefont {I.~A.}\ \bibnamefont {Walmsley}}, \ and\ \bibinfo {author}
  {\bibfnamefont {M.~G.}\ \bibnamefont {Raymer}},\ }\href@noop {} {\bibfield
  {journal} {\bibinfo  {journal} {arXiv preprint quant-ph/0611019}\ } (\bibinfo
  {year} {2006})}\BibitemShut {NoStop}%
\bibitem [{\citenamefont {Mosley}\ \emph {et~al.}(2008)\citenamefont {Mosley},
  \citenamefont {Lundeen}, \citenamefont {Smith},\ and\ \citenamefont
  {Walmsley}}]{mosley2008conditional}%
  \BibitemOpen
  \bibfield  {author} {\bibinfo {author} {\bibfnamefont {P.~J.}\ \bibnamefont
  {Mosley}}, \bibinfo {author} {\bibfnamefont {J.~S.}\ \bibnamefont {Lundeen}},
  \bibinfo {author} {\bibfnamefont {B.~J.}\ \bibnamefont {Smith}}, \ and\
  \bibinfo {author} {\bibfnamefont {I.~A.}\ \bibnamefont {Walmsley}},\
  }\href@noop {} {\bibfield  {journal} {\bibinfo  {journal} {New Journal of
  Physics}\ }\textbf {\bibinfo {volume} {10}},\ \bibinfo {pages} {093011}
  (\bibinfo {year} {2008})}\BibitemShut {NoStop}%
\bibitem [{\citenamefont {Bra{\'n}czyk}\ \emph {et~al.}(2011)\citenamefont
  {Bra{\'n}czyk}, \citenamefont {Fedrizzi}, \citenamefont {Stace},
  \citenamefont {Ralph},\ and\ \citenamefont {White}}]{branczyk2011engineered}%
  \BibitemOpen
  \bibfield  {author} {\bibinfo {author} {\bibfnamefont {A.~M.}\ \bibnamefont
  {Bra{\'n}czyk}}, \bibinfo {author} {\bibfnamefont {A.}~\bibnamefont
  {Fedrizzi}}, \bibinfo {author} {\bibfnamefont {T.~M.}\ \bibnamefont {Stace}},
  \bibinfo {author} {\bibfnamefont {T.~C.}\ \bibnamefont {Ralph}}, \ and\
  \bibinfo {author} {\bibfnamefont {A.~G.}\ \bibnamefont {White}},\ }\href@noop
  {} {\bibfield  {journal} {\bibinfo  {journal} {Optics Express}\ }\textbf
  {\bibinfo {volume} {19}},\ \bibinfo {pages} {55} (\bibinfo {year}
  {2011})}\BibitemShut {NoStop}%
\bibitem [{\citenamefont {Dixon}\ \emph {et~al.}(2013)\citenamefont {Dixon},
  \citenamefont {Shapiro},\ and\ \citenamefont {Wong}}]{dixon2013spectral}%
  \BibitemOpen
  \bibfield  {author} {\bibinfo {author} {\bibfnamefont {P.~B.}\ \bibnamefont
  {Dixon}}, \bibinfo {author} {\bibfnamefont {J.~H.}\ \bibnamefont {Shapiro}},
  \ and\ \bibinfo {author} {\bibfnamefont {F.~N.}\ \bibnamefont {Wong}},\
  }\href@noop {} {\bibfield  {journal} {\bibinfo  {journal} {Optics Express}\
  }\textbf {\bibinfo {volume} {21}},\ \bibinfo {pages} {5879} (\bibinfo {year}
  {2013})}\BibitemShut {NoStop}%
\bibitem [{\citenamefont {Dosseva}\ \emph {et~al.}(2016)\citenamefont
  {Dosseva}, \citenamefont {Cincio},\ and\ \citenamefont
  {Bra{\'n}czyk}}]{dosseva2016shaping}%
  \BibitemOpen
  \bibfield  {author} {\bibinfo {author} {\bibfnamefont {A.}~\bibnamefont
  {Dosseva}}, \bibinfo {author} {\bibfnamefont {{\L}.}~\bibnamefont {Cincio}},
  \ and\ \bibinfo {author} {\bibfnamefont {A.~M.}\ \bibnamefont
  {Bra{\'n}czyk}},\ }\href@noop {} {\bibfield  {journal} {\bibinfo  {journal}
  {Physical Review A}\ }\textbf {\bibinfo {volume} {93}},\ \bibinfo {pages}
  {013801} (\bibinfo {year} {2016})}\BibitemShut {NoStop}%
\bibitem [{\citenamefont {Tambasco}\ \emph {et~al.}(2016)\citenamefont
  {Tambasco}, \citenamefont {Boes}, \citenamefont {Helt}, \citenamefont
  {Steel},\ and\ \citenamefont {Mitchell}}]{tambasco2016domain}%
  \BibitemOpen
  \bibfield  {author} {\bibinfo {author} {\bibfnamefont {J.}~\bibnamefont
  {Tambasco}}, \bibinfo {author} {\bibfnamefont {A.}~\bibnamefont {Boes}},
  \bibinfo {author} {\bibfnamefont {L.}~\bibnamefont {Helt}}, \bibinfo {author}
  {\bibfnamefont {M.}~\bibnamefont {Steel}}, \ and\ \bibinfo {author}
  {\bibfnamefont {A.}~\bibnamefont {Mitchell}},\ }\href@noop {} {\bibfield
  {journal} {\bibinfo  {journal} {Optics Express}\ }\textbf {\bibinfo {volume}
  {24}},\ \bibinfo {pages} {19616} (\bibinfo {year} {2016})}\BibitemShut
  {NoStop}%
\bibitem [{\citenamefont {Graffitti}\ \emph {et~al.}(2017)\citenamefont
  {Graffitti}, \citenamefont {Kundys}, \citenamefont {Reid}, \citenamefont
  {Bra{\'n}czyk},\ and\ \citenamefont {Fedrizzi}}]{graffitti2017pure}%
  \BibitemOpen
  \bibfield  {author} {\bibinfo {author} {\bibfnamefont {F.}~\bibnamefont
  {Graffitti}}, \bibinfo {author} {\bibfnamefont {D.}~\bibnamefont {Kundys}},
  \bibinfo {author} {\bibfnamefont {D.~T.}\ \bibnamefont {Reid}}, \bibinfo
  {author} {\bibfnamefont {A.~M.}\ \bibnamefont {Bra{\'n}czyk}}, \ and\
  \bibinfo {author} {\bibfnamefont {A.}~\bibnamefont {Fedrizzi}},\ }\href
  {http://stacks.iop.org/2058-9565/2/i=3/a=035001} {\bibfield  {journal}
  {\bibinfo  {journal} {Quantum Science and Technology}\ }\textbf {\bibinfo
  {volume} {2}},\ \bibinfo {pages} {035001} (\bibinfo {year}
  {2017})}\BibitemShut {NoStop}%
\bibitem [{\citenamefont {Chen}\ \emph {et~al.}(2017)\citenamefont {Chen},
  \citenamefont {Bo}, \citenamefont {Niu}, \citenamefont {Xu}, \citenamefont
  {Zhang}, \citenamefont {Shapiro},\ and\ \citenamefont
  {Wong}}]{chen2017efficient}%
  \BibitemOpen
  \bibfield  {author} {\bibinfo {author} {\bibfnamefont {C.}~\bibnamefont
  {Chen}}, \bibinfo {author} {\bibfnamefont {C.}~\bibnamefont {Bo}}, \bibinfo
  {author} {\bibfnamefont {M.~Y.}\ \bibnamefont {Niu}}, \bibinfo {author}
  {\bibfnamefont {F.}~\bibnamefont {Xu}}, \bibinfo {author} {\bibfnamefont
  {Z.}~\bibnamefont {Zhang}}, \bibinfo {author} {\bibfnamefont {J.~H.}\
  \bibnamefont {Shapiro}}, \ and\ \bibinfo {author} {\bibfnamefont {F.~N.}\
  \bibnamefont {Wong}},\ }\href@noop {} {\bibfield  {journal} {\bibinfo
  {journal} {Optics Express}\ }\textbf {\bibinfo {volume} {25}},\ \bibinfo
  {pages} {7300} (\bibinfo {year} {2017})}\BibitemShut {NoStop}%
\bibitem [{\citenamefont {Shalm}\ \emph {et~al.}(2015)\citenamefont {Shalm},
  \citenamefont {Meyer-Scott}, \citenamefont {Christensen}, \citenamefont
  {Bierhorst}, \citenamefont {Wayne}, \citenamefont {Stevens}, \citenamefont
  {Gerrits}, \citenamefont {Glancy}, \citenamefont {Hamel}, \citenamefont
  {Allman} \emph {et~al.}}]{shalm2015strong}%
  \BibitemOpen
  \bibfield  {author} {\bibinfo {author} {\bibfnamefont {L.~K.}\ \bibnamefont
  {Shalm}}, \bibinfo {author} {\bibfnamefont {E.}~\bibnamefont {Meyer-Scott}},
  \bibinfo {author} {\bibfnamefont {B.~G.}\ \bibnamefont {Christensen}},
  \bibinfo {author} {\bibfnamefont {P.}~\bibnamefont {Bierhorst}}, \bibinfo
  {author} {\bibfnamefont {M.~A.}\ \bibnamefont {Wayne}}, \bibinfo {author}
  {\bibfnamefont {M.~J.}\ \bibnamefont {Stevens}}, \bibinfo {author}
  {\bibfnamefont {T.}~\bibnamefont {Gerrits}}, \bibinfo {author} {\bibfnamefont
  {S.}~\bibnamefont {Glancy}}, \bibinfo {author} {\bibfnamefont {D.~R.}\
  \bibnamefont {Hamel}}, \bibinfo {author} {\bibfnamefont {M.~S.}\ \bibnamefont
  {Allman}},  \emph {et~al.},\ }\href@noop {} {\bibfield  {journal} {\bibinfo
  {journal} {Physical Review Letters}\ }\textbf {\bibinfo {volume} {115}},\
  \bibinfo {pages} {250402} (\bibinfo {year} {2015})}\BibitemShut {NoStop}%
\bibitem [{\citenamefont {Fradkin}\ \emph {et~al.}(1999)\citenamefont
  {Fradkin}, \citenamefont {Arie}, \citenamefont {Skliar},\ and\ \citenamefont
  {Rosenman}}]{fradkin1999tunable}%
  \BibitemOpen
  \bibfield  {author} {\bibinfo {author} {\bibfnamefont {K.}~\bibnamefont
  {Fradkin}}, \bibinfo {author} {\bibfnamefont {A.}~\bibnamefont {Arie}},
  \bibinfo {author} {\bibfnamefont {A.}~\bibnamefont {Skliar}}, \ and\ \bibinfo
  {author} {\bibfnamefont {G.}~\bibnamefont {Rosenman}},\ }\href@noop {}
  {\bibfield  {journal} {\bibinfo  {journal} {Applied Physics Letters}\
  }\textbf {\bibinfo {volume} {74}},\ \bibinfo {pages} {914} (\bibinfo {year}
  {1999})}\BibitemShut {NoStop}%
\bibitem [{\citenamefont {Emanueli}\ and\ \citenamefont
  {Arie}(2003)}]{emanueli2003temperature}%
  \BibitemOpen
  \bibfield  {author} {\bibinfo {author} {\bibfnamefont {S.}~\bibnamefont
  {Emanueli}}\ and\ \bibinfo {author} {\bibfnamefont {A.}~\bibnamefont
  {Arie}},\ }\href@noop {} {\bibfield  {journal} {\bibinfo  {journal} {Applied
  Optics}\ }\textbf {\bibinfo {volume} {42}},\ \bibinfo {pages} {6661}
  (\bibinfo {year} {2003})}\BibitemShut {NoStop}%
\bibitem [{\citenamefont {K{\"o}nig}\ and\ \citenamefont
  {Wong}(2004)}]{konig2004extended}%
  \BibitemOpen
  \bibfield  {author} {\bibinfo {author} {\bibfnamefont {F.}~\bibnamefont
  {K{\"o}nig}}\ and\ \bibinfo {author} {\bibfnamefont {F.~N.}\ \bibnamefont
  {Wong}},\ }\href@noop {} {\bibfield  {journal} {\bibinfo  {journal} {Applied
  Physics Letters}\ }\textbf {\bibinfo {volume} {84}},\ \bibinfo {pages} {1644}
  (\bibinfo {year} {2004})}\BibitemShut {NoStop}%
\bibitem [{\citenamefont {Laudenbach}\ \emph {et~al.}(2016)\citenamefont
  {Laudenbach}, \citenamefont {H{\"u}bel}, \citenamefont {Hentschel},
  \citenamefont {Walther},\ and\ \citenamefont
  {Poppe}}]{laudenbach2016modelling}%
  \BibitemOpen
  \bibfield  {author} {\bibinfo {author} {\bibfnamefont {F.}~\bibnamefont
  {Laudenbach}}, \bibinfo {author} {\bibfnamefont {H.}~\bibnamefont
  {H{\"u}bel}}, \bibinfo {author} {\bibfnamefont {M.}~\bibnamefont
  {Hentschel}}, \bibinfo {author} {\bibfnamefont {P.}~\bibnamefont {Walther}},
  \ and\ \bibinfo {author} {\bibfnamefont {A.}~\bibnamefont {Poppe}},\
  }\href@noop {} {\bibfield  {journal} {\bibinfo  {journal} {Optics Express}\
  }\textbf {\bibinfo {volume} {24}},\ \bibinfo {pages} {2712} (\bibinfo {year}
  {2016})}\BibitemShut {NoStop}%
\bibitem [{\citenamefont {Avenhaus}\ \emph {et~al.}(2009)\citenamefont
  {Avenhaus}, \citenamefont {Eckstein}, \citenamefont {Mosley},\ and\
  \citenamefont {Silberhorn}}]{avenhaus2009fiber}%
  \BibitemOpen
  \bibfield  {author} {\bibinfo {author} {\bibfnamefont {M.}~\bibnamefont
  {Avenhaus}}, \bibinfo {author} {\bibfnamefont {A.}~\bibnamefont {Eckstein}},
  \bibinfo {author} {\bibfnamefont {P.~J.}\ \bibnamefont {Mosley}}, \ and\
  \bibinfo {author} {\bibfnamefont {C.}~\bibnamefont {Silberhorn}},\
  }\href@noop {} {\bibfield  {journal} {\bibinfo  {journal} {Optics Letters}\
  }\textbf {\bibinfo {volume} {34}},\ \bibinfo {pages} {2873} (\bibinfo {year}
  {2009})}\BibitemShut {NoStop}%
\bibitem [{\citenamefont {Weston}\ \emph {et~al.}(2016)\citenamefont {Weston},
  \citenamefont {Chrzanowski}, \citenamefont {Wollmann}, \citenamefont
  {Boston}, \citenamefont {Ho}, \citenamefont {Shalm}, \citenamefont {Verma},
  \citenamefont {Allman}, \citenamefont {Nam}, \citenamefont {Patel} \emph
  {et~al.}}]{weston2016efficient}%
  \BibitemOpen
  \bibfield  {author} {\bibinfo {author} {\bibfnamefont {M.~M.}\ \bibnamefont
  {Weston}}, \bibinfo {author} {\bibfnamefont {H.~M.}\ \bibnamefont
  {Chrzanowski}}, \bibinfo {author} {\bibfnamefont {S.}~\bibnamefont
  {Wollmann}}, \bibinfo {author} {\bibfnamefont {A.}~\bibnamefont {Boston}},
  \bibinfo {author} {\bibfnamefont {J.}~\bibnamefont {Ho}}, \bibinfo {author}
  {\bibfnamefont {L.~K.}\ \bibnamefont {Shalm}}, \bibinfo {author}
  {\bibfnamefont {V.~B.}\ \bibnamefont {Verma}}, \bibinfo {author}
  {\bibfnamefont {M.~S.}\ \bibnamefont {Allman}}, \bibinfo {author}
  {\bibfnamefont {S.~W.}\ \bibnamefont {Nam}}, \bibinfo {author} {\bibfnamefont
  {R.~B.}\ \bibnamefont {Patel}},  \emph {et~al.},\ }\href@noop {} {\bibfield
  {journal} {\bibinfo  {journal} {Optics Express}\ }\textbf {\bibinfo {volume}
  {24}},\ \bibinfo {pages} {10869} (\bibinfo {year} {2016})}\BibitemShut
  {NoStop}%
\bibitem [{\citenamefont {Liscidini}\ and\ \citenamefont
  {Sipe}(2013)}]{liscidini2013stimulated}%
  \BibitemOpen
  \bibfield  {author} {\bibinfo {author} {\bibfnamefont {M.}~\bibnamefont
  {Liscidini}}\ and\ \bibinfo {author} {\bibfnamefont {J.}~\bibnamefont
  {Sipe}},\ }\href@noop {} {\bibfield  {journal} {\bibinfo  {journal} {Physical
  Review Letters}\ }\textbf {\bibinfo {volume} {111}},\ \bibinfo {pages}
  {193602} (\bibinfo {year} {2013})}\BibitemShut {NoStop}%
\bibitem [{\citenamefont {Eckstein}\ \emph {et~al.}(2014)\citenamefont
  {Eckstein}, \citenamefont {Boucher}, \citenamefont {Lema{\^\i}tre},
  \citenamefont {Filloux}, \citenamefont {Favero}, \citenamefont {Leo},
  \citenamefont {Sipe}, \citenamefont {Liscidini},\ and\ \citenamefont
  {Ducci}}]{eckstein2014high}%
  \BibitemOpen
  \bibfield  {author} {\bibinfo {author} {\bibfnamefont {A.}~\bibnamefont
  {Eckstein}}, \bibinfo {author} {\bibfnamefont {G.}~\bibnamefont {Boucher}},
  \bibinfo {author} {\bibfnamefont {A.}~\bibnamefont {Lema{\^\i}tre}}, \bibinfo
  {author} {\bibfnamefont {P.}~\bibnamefont {Filloux}}, \bibinfo {author}
  {\bibfnamefont {I.}~\bibnamefont {Favero}}, \bibinfo {author} {\bibfnamefont
  {G.}~\bibnamefont {Leo}}, \bibinfo {author} {\bibfnamefont {J.~E.}\
  \bibnamefont {Sipe}}, \bibinfo {author} {\bibfnamefont {M.}~\bibnamefont
  {Liscidini}}, \ and\ \bibinfo {author} {\bibfnamefont {S.}~\bibnamefont
  {Ducci}},\ }\href@noop {} {\bibfield  {journal} {\bibinfo  {journal} {Laser
  \& Photonics Reviews}\ }\textbf {\bibinfo {volume} {8}} (\bibinfo {year}
  {2014})}\BibitemShut {NoStop}%
\bibitem [{\citenamefont {Law}\ \emph {et~al.}(2000)\citenamefont {Law},
  \citenamefont {Walmsley},\ and\ \citenamefont {Eberly}}]{law2000continuous}%
  \BibitemOpen
  \bibfield  {author} {\bibinfo {author} {\bibfnamefont {C.}~\bibnamefont
  {Law}}, \bibinfo {author} {\bibfnamefont {I.}~\bibnamefont {Walmsley}}, \
  and\ \bibinfo {author} {\bibfnamefont {J.}~\bibnamefont {Eberly}},\
  }\href@noop {} {\bibfield  {journal} {\bibinfo  {journal} {Physical Review
  Letters}\ }\textbf {\bibinfo {volume} {84}},\ \bibinfo {pages} {5304}
  (\bibinfo {year} {2000})}\BibitemShut {NoStop}%
\bibitem [{\citenamefont {Giovannetti}\ \emph {et~al.}(2002)\citenamefont
  {Giovannetti}, \citenamefont {Maccone}, \citenamefont {Shapiro},\ and\
  \citenamefont {Wong}}]{giovannetti2002extended}%
  \BibitemOpen
  \bibfield  {author} {\bibinfo {author} {\bibfnamefont {V.}~\bibnamefont
  {Giovannetti}}, \bibinfo {author} {\bibfnamefont {L.}~\bibnamefont
  {Maccone}}, \bibinfo {author} {\bibfnamefont {J.~H.}\ \bibnamefont
  {Shapiro}}, \ and\ \bibinfo {author} {\bibfnamefont {F.~N.}\ \bibnamefont
  {Wong}},\ }\href@noop {} {\bibfield  {journal} {\bibinfo  {journal} {Physical
  Review A}\ }\textbf {\bibinfo {volume} {66}},\ \bibinfo {pages} {043813}
  (\bibinfo {year} {2002})}\BibitemShut {NoStop}%
\bibitem [{\citenamefont {Bra{\'n}czyk}(2017)}]{branczyk2017hong}%
  \BibitemOpen
  \bibfield  {author} {\bibinfo {author} {\bibfnamefont {A.~M.}\ \bibnamefont
  {Bra{\'n}czyk}},\ }\href@noop {} {\bibfield  {journal} {\bibinfo  {journal}
  {arXiv preprint arXiv:1711.00080}\ } (\bibinfo {year} {2017})}\BibitemShut
  {NoStop}%
\bibitem [{\citenamefont {Kok}\ \emph {et~al.}(2007)\citenamefont {Kok},
  \citenamefont {Munro}, \citenamefont {Nemoto}, \citenamefont {Ralph},
  \citenamefont {Dowling},\ and\ \citenamefont {Milburn}}]{kok2007linear}%
  \BibitemOpen
  \bibfield  {author} {\bibinfo {author} {\bibfnamefont {P.}~\bibnamefont
  {Kok}}, \bibinfo {author} {\bibfnamefont {W.~J.}\ \bibnamefont {Munro}},
  \bibinfo {author} {\bibfnamefont {K.}~\bibnamefont {Nemoto}}, \bibinfo
  {author} {\bibfnamefont {T.~C.}\ \bibnamefont {Ralph}}, \bibinfo {author}
  {\bibfnamefont {J.~P.}\ \bibnamefont {Dowling}}, \ and\ \bibinfo {author}
  {\bibfnamefont {G.~J.}\ \bibnamefont {Milburn}},\ }\href@noop {} {\bibfield
  {journal} {\bibinfo  {journal} {Reviews of Modern Physics}\ }\textbf
  {\bibinfo {volume} {79}},\ \bibinfo {pages} {135} (\bibinfo {year}
  {2007})}\BibitemShut {NoStop}%
\bibitem [{\citenamefont {Barbieri}\ \emph {et~al.}(2017)\citenamefont
  {Barbieri}, \citenamefont {Roccia}, \citenamefont {Mancino}, \citenamefont
  {Sbroscia}, \citenamefont {Gianani},\ and\ \citenamefont
  {Sciarrino}}]{barbieri2017hong}%
  \BibitemOpen
  \bibfield  {author} {\bibinfo {author} {\bibfnamefont {M.}~\bibnamefont
  {Barbieri}}, \bibinfo {author} {\bibfnamefont {E.}~\bibnamefont {Roccia}},
  \bibinfo {author} {\bibfnamefont {L.}~\bibnamefont {Mancino}}, \bibinfo
  {author} {\bibfnamefont {M.}~\bibnamefont {Sbroscia}}, \bibinfo {author}
  {\bibfnamefont {I.}~\bibnamefont {Gianani}}, \ and\ \bibinfo {author}
  {\bibfnamefont {F.}~\bibnamefont {Sciarrino}},\ }\href@noop {} {\bibfield
  {journal} {\bibinfo  {journal} {Scientific Reports}\ }\textbf {\bibinfo
  {volume} {7}} (\bibinfo {year} {2017})}\BibitemShut {NoStop}%
\bibitem [{\citenamefont {Fedrizzi}\ \emph {et~al.}(2009)\citenamefont
  {Fedrizzi}, \citenamefont {Herbst}, \citenamefont {Aspelmeyer}, \citenamefont
  {Barbieri}, \citenamefont {Jennewein},\ and\ \citenamefont
  {Zeilinger}}]{fedrizzi2009anti}%
  \BibitemOpen
  \bibfield  {author} {\bibinfo {author} {\bibfnamefont {A.}~\bibnamefont
  {Fedrizzi}}, \bibinfo {author} {\bibfnamefont {T.}~\bibnamefont {Herbst}},
  \bibinfo {author} {\bibfnamefont {M.}~\bibnamefont {Aspelmeyer}}, \bibinfo
  {author} {\bibfnamefont {M.}~\bibnamefont {Barbieri}}, \bibinfo {author}
  {\bibfnamefont {T.}~\bibnamefont {Jennewein}}, \ and\ \bibinfo {author}
  {\bibfnamefont {A.}~\bibnamefont {Zeilinger}},\ }\href@noop {} {\bibfield
  {journal} {\bibinfo  {journal} {New Journal of Physics}\ }\textbf {\bibinfo
  {volume} {11}},\ \bibinfo {pages} {103052} (\bibinfo {year}
  {2009})}\BibitemShut {NoStop}%
\bibitem [{\citenamefont {Fedrizzi}\ \emph {et~al.}(2007)\citenamefont
  {Fedrizzi}, \citenamefont {Herbst}, \citenamefont {Poppe}, \citenamefont
  {Jennewein},\ and\ \citenamefont {Zeilinger}}]{fedrizzi2007wavelength}%
  \BibitemOpen
  \bibfield  {author} {\bibinfo {author} {\bibfnamefont {A.}~\bibnamefont
  {Fedrizzi}}, \bibinfo {author} {\bibfnamefont {T.}~\bibnamefont {Herbst}},
  \bibinfo {author} {\bibfnamefont {A.}~\bibnamefont {Poppe}}, \bibinfo
  {author} {\bibfnamefont {T.}~\bibnamefont {Jennewein}}, \ and\ \bibinfo
  {author} {\bibfnamefont {A.}~\bibnamefont {Zeilinger}},\ }\href@noop {}
  {\bibfield  {journal} {\bibinfo  {journal} {Optics Express}\ }\textbf
  {\bibinfo {volume} {15}},\ \bibinfo {pages} {15377} (\bibinfo {year}
  {2007})}\BibitemShut {NoStop}%
\bibitem [{\citenamefont {Fejer}\ \emph {et~al.}(1992)\citenamefont {Fejer},
  \citenamefont {Magel}, \citenamefont {Jundt},\ and\ \citenamefont
  {Byer}}]{fejer1992quasi}%
  \BibitemOpen
  \bibfield  {author} {\bibinfo {author} {\bibfnamefont {M.~M.}\ \bibnamefont
  {Fejer}}, \bibinfo {author} {\bibfnamefont {G.}~\bibnamefont {Magel}},
  \bibinfo {author} {\bibfnamefont {D.~H.}\ \bibnamefont {Jundt}}, \ and\
  \bibinfo {author} {\bibfnamefont {R.~L.}\ \bibnamefont {Byer}},\ }\href@noop
  {} {\bibfield  {journal} {\bibinfo  {journal} {IEEE Journal of Quantum
  Electronics}\ }\textbf {\bibinfo {volume} {28}},\ \bibinfo {pages} {2631}
  (\bibinfo {year} {1992})}\BibitemShut {NoStop}%
\bibitem [{\citenamefont {Pelc}\ \emph {et~al.}(2010)\citenamefont {Pelc},
  \citenamefont {Langrock}, \citenamefont {Zhang},\ and\ \citenamefont
  {Fejer}}]{pelc2010influence}%
  \BibitemOpen
  \bibfield  {author} {\bibinfo {author} {\bibfnamefont {J.~S.}\ \bibnamefont
  {Pelc}}, \bibinfo {author} {\bibfnamefont {C.}~\bibnamefont {Langrock}},
  \bibinfo {author} {\bibfnamefont {Q.}~\bibnamefont {Zhang}}, \ and\ \bibinfo
  {author} {\bibfnamefont {M.~M.}\ \bibnamefont {Fejer}},\ }\href@noop {}
  {\bibfield  {journal} {\bibinfo  {journal} {Optics Letters}\ }\textbf
  {\bibinfo {volume} {35}},\ \bibinfo {pages} {2804} (\bibinfo {year}
  {2010})}\BibitemShut {NoStop}%
\bibitem [{\citenamefont {Phillips}\ \emph {et~al.}(2013)\citenamefont
  {Phillips}, \citenamefont {Pelc},\ and\ \citenamefont
  {Fejer}}]{phillips2013parametric}%
  \BibitemOpen
  \bibfield  {author} {\bibinfo {author} {\bibfnamefont {C.}~\bibnamefont
  {Phillips}}, \bibinfo {author} {\bibfnamefont {J.}~\bibnamefont {Pelc}}, \
  and\ \bibinfo {author} {\bibfnamefont {M.}~\bibnamefont {Fejer}},\
  }\href@noop {} {\bibfield  {journal} {\bibinfo  {journal} {JOSA B}\ }\textbf
  {\bibinfo {volume} {30}},\ \bibinfo {pages} {982} (\bibinfo {year}
  {2013})}\BibitemShut {NoStop}%
\bibitem [{\citenamefont {Graffitti~et al.}(2018)}]{graffitti2018poled}%
  \BibitemOpen
  \bibfield  {author} {\bibinfo {author} {\bibfnamefont {F.}~\bibnamefont
  {Graffitti~et al.}},\ }\href@noop {} {\bibfield  {journal} {\bibinfo
  {journal} {Under preparation}\ } (\bibinfo {year} {2018})}\BibitemShut
  {NoStop}%
\bibitem [{\citenamefont {Hong}\ \emph {et~al.}(1987)\citenamefont {Hong},
  \citenamefont {Ou},\ and\ \citenamefont {Mandel}}]{hong1987measurement}%
  \BibitemOpen
  \bibfield  {author} {\bibinfo {author} {\bibfnamefont {C.}~\bibnamefont
  {Hong}}, \bibinfo {author} {\bibfnamefont {Z.}~\bibnamefont {Ou}}, \ and\
  \bibinfo {author} {\bibfnamefont {L.}~\bibnamefont {Mandel}},\ }\href@noop {}
  {\bibfield  {journal} {\bibinfo  {journal} {Physical Review Letters}\
  }\textbf {\bibinfo {volume} {59}},\ \bibinfo {pages} {2044} (\bibinfo {year}
  {1987})}\BibitemShut {NoStop}%
\end{thebibliography}%

\clearpage

\appendix {\textbf{\noindent Appendix A: Visibility at low power}}

\noindent 
The two-photon interference visibility is defined as:
\begin{equation}
V=\frac{N_{\text{max}}-N_{\text{min}}}{N_{\text{max}}} \, ,
\label{HOMexperiment}
\end{equation}
where $N_{\text{max}}$ ($N_{\text{min}}$) is the maximum (minimum) number of coincidences recorded during the two-photon interference scan, in which the arrival time of the photons at the BS from its two inputs is varied by mean of a delay line.
In a standard two-photon interference experiment, $N_{\text{max}}$ corresponds to counts recorded when the photons arrive at the BS at perfectly distinguishable times: no quantum interference occurs in such a case, as the photons are distinguishable in the temporal degree of freedom. We will refer to this configuration as ``outside the dip''.
$N_{\text{min}}$ is instead the number of coincidences recorded when the photons arrive simultaneously at the BS: in this case quantum interference effects arise and, when two and only two indistinguishable photons enter the BS we have perfect two-photon bunching \cite{hong1987measurement}. We will refer to this configuration as ``in the dip''.

Equation \eqref{HOMexperiment} can be written in terms of the probabilities of having a coincidence ``outside the dip'' ($p_{cc}^\text{out}$) and ``in the dip''  ($p_{cc}^\text{in}$):
\begin{equation}
V(\lambda)=\frac{p_{cc}^\text{out} (\lambda) -p_{cc}^\text{in} (\lambda) }{p_{cc}^\text{out} (\lambda) } \, ,
\label{HOMprob}
\end{equation}
In order to find the visibility dependence on the $\lambda$ parameter (which translates to its dependence on the pump power, as $\lambda = \sqrt{P \tau}$ \cite{kok2007linear,broome2011reducing}), we need to consider the action of the BS on the higher order terms of the PDC states.

To this aim, we consider the transformations introduced by the BS on the creation operators ($a^\dagger$ and $b^\dagger$) of the input photons \cite{branczyk2017hong}:
\begin{equation}
\begin{split}
&a^\dagger \xrightarrow{\text{BS}} \frac{c^\dagger + i\ d^\dagger}{\sqrt{2}}\\
&b^\dagger \xrightarrow{\text{BS}}\frac{i\ c^\dagger + d^\dagger}{\sqrt{2}}\, .
\end{split}
\label{bs}
\end{equation}
We also write the PDC state in the Fock space in terms of the creation operators $a^\dagger$ and $b^\dagger$ :
\begin{equation}
\begin{split}
\ket{\psi_{\text{PDC}}}&=\sqrt{1-\abs{\lambda}^2} \sum_{n=0}^\infty \lambda^n \ket{n}_s \ket{n}_i\\
&= \sqrt{1-\abs{\lambda}^2} \sum_{n=0}^\infty \lambda^n \frac{\left(a^\dagger b^\dagger\right)^n}{n!} \ket{0}_s \ket{0}_i\, .
\end{split}
\label{PDCoper}
\end{equation}

In this framework, the probability of having coincidences after the BS corresponds to the amplitudes squared of the $\ket{n>0}_c\ket{m>0}_d$ terms in the BS-output Fock-state,
i.e. all the terms where both the $c^\dagger$ and $d^\dagger$ operators occur.
This probability is calculated as $\sum_{n,m>0} \abs{\braket{n,m|\psi_{\text{BS}}}}^2$, where $\ket{\psi_{\text{BS}}}$ is the state after the BS.
In the ``outside the dip'' case, an additional term has to be taken into account for a correct estimation of $p_{cc}^\text{out}$, as discussed in the next section.

\subsection*{Outside the dip}
We can write the state after the BS by mean of the transformations in \eqref{bs} applied to one PDC state (\eqref{PDCoper}) (in the case of signal-idler interference):
\begin{equation}
\begin{split}
\ket{\psi_{\text{BS}}} &= \sqrt{1-\abs{\lambda}^2} \sum_{n=0}^\infty \lambda^n \frac{\left(c^\dagger_1 + i\  d^\dagger_1\right)^n \left(i\ c^\dagger_2 +  d^\dagger_2\right)^n}{2^n n!} \ket{0}_s \ket{0}_i,
\end{split}
\label{outSigIdl}
\end{equation}
or to two heralded PDC photons:
\begin{equation}
\begin{split}
\ket{\psi_{\text{BS}}} = \left(1-\abs{\lambda}^2\right) &\times\\
\times\sum_{n,m=0}^\infty \lambda^{n+m} &\frac{\left(c^\dagger_1 + i\  d^\dagger_1\right)^n \left(i\ c^\dagger_2 +  d^\dagger_2\right)^m}{2^{n+m} n! m!} \ket{0}_s \ket{0}_i\, ,
\end{split}
\label{outInd}
\end{equation}
where we have used the subscripts $1$ and $2$ to indicate the different arrival time to the BS of the photons $a^\dagger_1$ and $b^\dagger_2$. For simplicity, we use the same $\lambda$ parameter for both the PDC processes.

We can then calculate the probability of having coincidences ``outside the dip''.
We assume a lossless setup, and that the detectors and the logic can't resolve the detection time of the photons:
this is reasonable for an actual experimental setup because the difference in the photons detection time is usually narrower than the counting logic resolution.
Under this assumptions, we calculate $p_{cc}^\text{out}$ as the sum of three terms:
\begin{enumerate}[label=(\roman*)]
\item $p_{cc}(a^\dagger_1) p_{cc}(b^\dagger_2)$
\item $(1-p_{cc}(a^\dagger_1)) p_{cc}(b^\dagger_2) + p_{cc}(a^\dagger_1) (1-p_{cc}(b^\dagger_1))$
\item $\frac{1}{2}(1-p_{cc}(a^\dagger_1)) (1-p_{cc}(b^\dagger_2))$
\end{enumerate}
The first term corresponds to the probability that photons from both the inputs of the BS ($a^\dagger_1$,$b^\dagger_2$)  lead to coincidences:  the detection logic clearly records this case as a coincidence event.
The second term corresponds to the probability that photons from one of the two inputs don't give coincidences but the others do, and vice versa: also in this case the detection logic records a coincidence event, as at least one photon for each BS output is detected in the logic time window.
The third additional term corresponds to the probability that none of the photons from the two inputs give coincidences, but photons from one input go to the opposite output respect to the photons of the other input: this occurs with probability 1/2, and the logic detects it as a coincidence event because it can't resolve the difference in the arrival time of the photons from the two BS outputs.

\subsection*{In the dip}
We can also calculate the probability of having coincidences ``in the dip'' by considering the state after the BS when the photons from the two inputs arrive simultaneously at the BS.
In the case of signal-idler interference, the output state reads:
\begin{equation}
\begin{split}
\ket{\psi_{\text{BS}}} &= \sqrt{1-\abs{\lambda}^2} \sum_{n=0}^\infty \lambda^n \frac{\left(c^\dagger + i\  d^\dagger\right)^n \left(i\ c^\dagger +  d^\dagger\right)^n}{2^n n!} \ket{0}_s \ket{0}_i\\
&=\sqrt{1-\abs{\lambda}^2} \sum_{n=0}^\infty \lambda^n\ i^n \frac{  \left(c^{\dagger 2} +  d^{\dagger 2}\right)^n }{2^n n!} \ket{0}_s \ket{0}_i\, ,
\end{split}
\label{inSigIdl}
\end{equation}
while in the heralded PDC photons case it reads:
\begin{equation}
\begin{split}
\ket{\psi_{\text{BS}}} = \left(1-\abs{\lambda}^2\right) &\times\\
\times \sum_{n,m=0}^\infty \lambda^{n+m} &\frac{\left(c^\dagger + i\  d^\dagger\right)^n \left(i\ c^\dagger +  d^\dagger\right)^m}{2^{n+m} n! m!} \ket{0}_s \ket{0}_i\, .
\end{split}
\label{inInd}
\end{equation}
In this configuration we don't have the subscripts $1$ and $2$ as in \eqref{outSigIdl} and \eqref{outInd} because
it's impossible to distinguish the arrival time of the photons at the BS.

\subsection*{Visibility up to the fifth PDC order}
We can finally find the two-photon interference visibility as a function of $\lambda$ up to a given PDC order. 
We calculate $p_{cc}^\text{out} (\lambda)$ according to the bullet points in the ``Outside the dip'' section , starting from the state in \eqref{outSigIdl} (or in \eqref{outInd}) and computing the corresponding probabilities.
We also calculate $p_{cc}^\text{in} (\lambda)$ by computing the coincidence probabilities corresponding to the state in \eqref{inSigIdl} (or in \eqref{inInd}). 
We can therefore plug these values in \eqref{HOMprob} in order to calculate $V(\lambda)$. 
\newpage
 
\begin{figure}[htb]\center
\includegraphics[width=1\columnwidth]{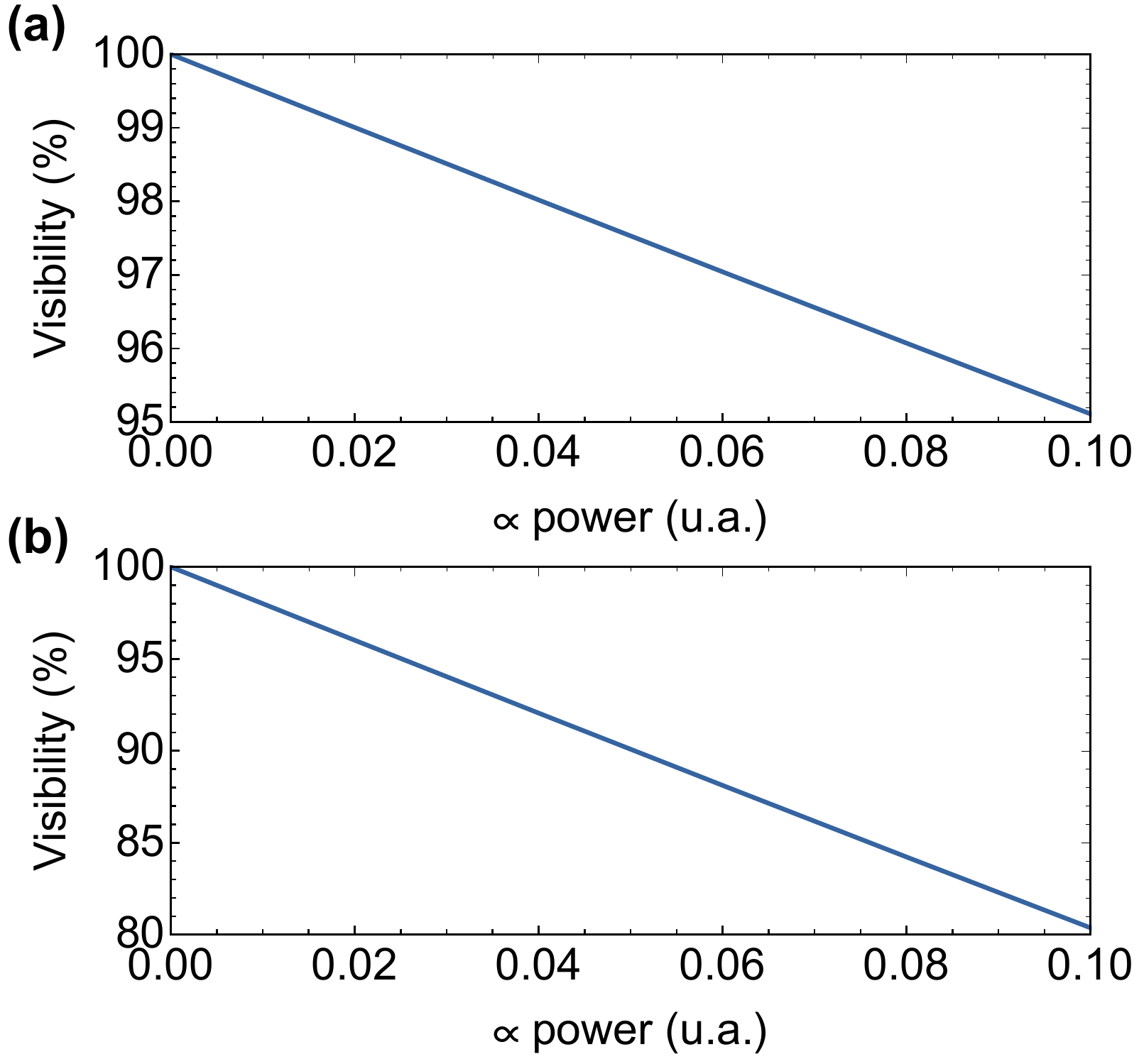}
\caption{
(a) Two-photon visibility as a function of the pump power for the signal-idler interference.
(b) Two-photon visibility as a function of the pump power for the heralded-photons interference.
}
\label{HOMvisibility}
\end{figure}

As an example, we show here the two-photon interference visibility up to the fifth order of the PDC state for the case of 
signal-idler interference (see fig. \ref{HOMvisibility} (a)) and heralded-photons interference (see fig. \ref{HOMvisibility} (b)).
Truncating to the fifth order of the PDC state is enough for a reliable estimation of $V(\lambda)$ in the chosen $\lambda$ range, as
the difference of the visibility when we expand the PDC state up to the fourth and the fifth order is less than \SI{0.02}{\percent} .
In both the signal-idler interference  and the heralded-photons interference, the visibility decreases approximately linearly as the power increases (at low $\lambda$ values). 

\end{document}